\pdfoutput=1
\documentclass{article}
\usepackage{algorithm}
\usepackage{footnote}
\usepackage{enumitem}
\usepackage{adjustbox}
\usepackage{textgreek}

\usepackage{fullpage}

\usepackage[utf8]{inputenc} 
\usepackage[T1]{fontenc}    
\usepackage{hyperref}       
\usepackage{url}            
\usepackage{booktabs}       
\usepackage{amsfonts}       
\usepackage{nicefrac}       
\usepackage{microtype}      
\usepackage{xcolor}
\usepackage{acronym}
\usepackage{subfigure}
\usepackage{graphicx,multicol}
\usepackage{algpseudocode}
\usepackage{mathrsfs}  
\usepackage{comment}
\usepackage{mathtools}
\usepackage{natbib}

\definecolor{NavyBlue}{RGB}{35,35,142}
\definecolor{RawSienna}{RGB}{199,97,20}
\hypersetup{
  colorlinks,%
  citecolor=NavyBlue,%
  filecolor=NavyBlue,%
  linkcolor=RawSienna,%
  urlcolor=NavyBlue
}

\DeclareMathOperator*{\argmin}{argmin}

\newcommand*{\argminl}{\argmin\limits}

\newcommand{\rr}{\mathbb{R}}
\def\RR{\mathbb{R}}
\newcommand{\1}[1]{\mathbf{1}\!\left[#1\right]}
\newcommand{\norm}[1]{\left\lVert #1\right\rVert}
\newcommand{\g}{\pmb{\gamma}}

\newcommand{\p}{\mathbf{p}}

\def\C{\mathbf{C}}

\newcommand{\w}{\mathbf{w}}

\newcommand{\x}{\mathbf{x}}
\def\r{\mathbf{r}}

\newcommand{\Bu}{\pmb{\upsilon}}

\newcommand{\z}{\mathbf{z}}

\def\PP{\mathbb{P}}
\def\clip{{\text{clip}}}
\def\ones{\mathbf{1}}
\def\oon{{\textstyle \frac{1}{n}}}
\newcommand\abs[1]{\left|#1\right|}

\newcommand{\task}[1]{{\fontfamily{cmr}\selectfont #1}}

\def\optirank{{\texttt{optirank}}}
\def\ANranklr{{\texttt{ANrank-lr}}}
\def\lr{{\texttt{lr}}}
\def\ranklr{{\texttt{rank-lr}}}
\def\rf{{\texttt{rf}}}
\def\SCN{{\texttt{SCN}}}
\definecolor{mygreen}{rgb}{0,0.5,0}

\def\sorlr{s_{\text{\texttt{or}}}}
\def\slr{s_{\text{\texttt{lr}}}}
\def\sAr{s_{\text{\texttt{Ar}}}}
\def\sothers{s_{\text{\texttt{others}}}}

\title{Optirank: classification for RNA-Seq data with optimal ranking reference genes}

\author{%
  Paola Malsot$^{1,}$\footnote{\texttt{paola.malsot@epfl.ch}}\,, Filipe Martins$^1$, Didier Trono$^1$, Guillaume Obozinski$^{2,}$\footnote{\texttt{guillaume.obozinski@epfl.ch}} \\[1.5mm]
  $^1$Ecole Polytechnique Fédérale de Lausanne\\ $^2$Swiss Data Science Center, EPFL \& ETH Zürich
}

\begin{document}

\maketitle

\begin{abstract}
Classification algorithms using RNA-Sequencing (RNA-Seq) data as input are used in a variety of biological applications. By nature, RNA-Seq data is subject to uncontrolled fluctuations both within and especially across datasets, which presents a major difficulty for a trained classifier to generalize to an external dataset. \\
Replacing raw gene counts with the rank of gene counts inside an observation has proven effective to mitigate this problem. However, the rank of a feature is by definition relative to all other features, including highly variable features that introduce noise in the ranking. \\
To address this problem and obtain more robust ranks, we propose a logistic regression model, \optirank, which learns simultaneously the parameters of the model and the genes to use as a \emph{reference set} in the ranking. \\
We show the effectiveness of this method on simulated data. We also consider real classification tasks, which present different kinds of distribution shifts between train and test data. Those tasks concern a variety of applications, such as cancer of unknown primary classification, identification of specific gene signatures, and determination of cell type in single-cell RNA-Seq datasets. On those real tasks, \optirank~performs at least as well as the vanilla logistic regression on classical ranks, while producing sparser solutions. \\
In addition, to increase the robustness against dataset shifts, we propose a multi-source learning scheme and demonstrate its effectiveness when used in combination with rank-based classifiers. 
\end{abstract}

\section{Introduction}

RNA-Sequencing provides a way to probe the state of cells and tissues, by measuring the level of expression of thousands of genes. Since its introduction, RNA-Seq data has been used in differential expression analysis to highlight genes that are differentially expressed in two contrasting conditions (stereotypically healthy versus diseased), pointing towards potentially actionable drug targets and molecular mechanisms. In this context, normalization of RNA-Seq data has been extensively studied: we provide in the subsequent section an overview of common methods. However there is still a lack of consensus on normalization for classification tasks, which is crucial given the recent emergence of machine-learning assisted diagnosis based on RNA-Seq data \citep[for instance][]{breast-subtyping,TOD-CUP,singlecellnet}.

In practice, among other normalization techniques also used in differential expression analysis, ranking normalization seems to have had particular success in combination with classification algorithms \citep{TOD-CUP,cell-cycle-rank}. Ranking normalization consists simply in replacing the raw read count of genes by their ranks amongst the read count of other genes for the same observation. \cite{rank-based-classifiers} show a consistent improvement of score when ranking normalization is used.

A potential weakness of ranking normalization is that the rank of an otherwise informative gene could be perturbed by genes whose expression fluctuate independently from the variable of interest. An obvious solution is to rank gene expressions only relative to a set of stable genes, which we call \emph{reference set}. The difficulty is, however, in choosing this set. With this motivation, and to solve binary classification problems based on robust and adaptive ranks, we propose \optirank, a logistic regression model based on ranks relative to a \emph{reference set}, where the latter is learned at the same time as the weights of the logistic model.

\subsection{Overview of Normalization Techniques}

Multiple factors alter the number of reads obtained for a gene beyond the number of corresponding RNA molecules in the biological sample, the quantity of interest. For instance, the preservation technique of a biological sample and its temperature influence the natural degradation process of RNA; the length and the GC-content of an RNA molecule will affect its reading rate. Normalization aims at obtaining a representation of the data invariant to those aforementioned nuisance factors. Ideally, the remaining variation after normalization should be attributable solely to the phenomenon of interest (stereotypically, disease versus normal). In this way, the analyst avoids the risk of confounding the effect of the variable of interest with the effect of the nuisance variables, and the learned model is applicable to an external dataset obtained with a different configuration of nuisance factors. \\
The simplest normalization, total count normalization, divides the raw read count on each gene by the total number of reads in the corresponding observation. However, highly expressed genes can induce a misleadingly low proportion of reads on other genes which are normally expressed. More refined techniques, such as TMM and DESeq counteract this artifact by using a more robust normalization factor. Quantile normalization matches the distribution of gene expressions to a reference distribution. Other methods correct the expression thanks to controls, such as housekeeping genes hypothesized to be constant across conditions or RNA spike-ins whose quantities are controlled during library preparation.\\
For the interested reader, \cite{review-normalization-1} and  \cite{review-normalization-2} evaluate these methods in differential expression analysis with real and simulated data. To our knowledge, there is no similar review on the use of RNA-Seq normalization as a pre-processing step prior to classification.

\section{Learning from ranks}

\subsection{Ranking with respect to a \emph{reference set}} \label{sec:ranking-ref-main}

\begin{figure}[h!]
\centering
\includegraphics[width=0.9\linewidth]{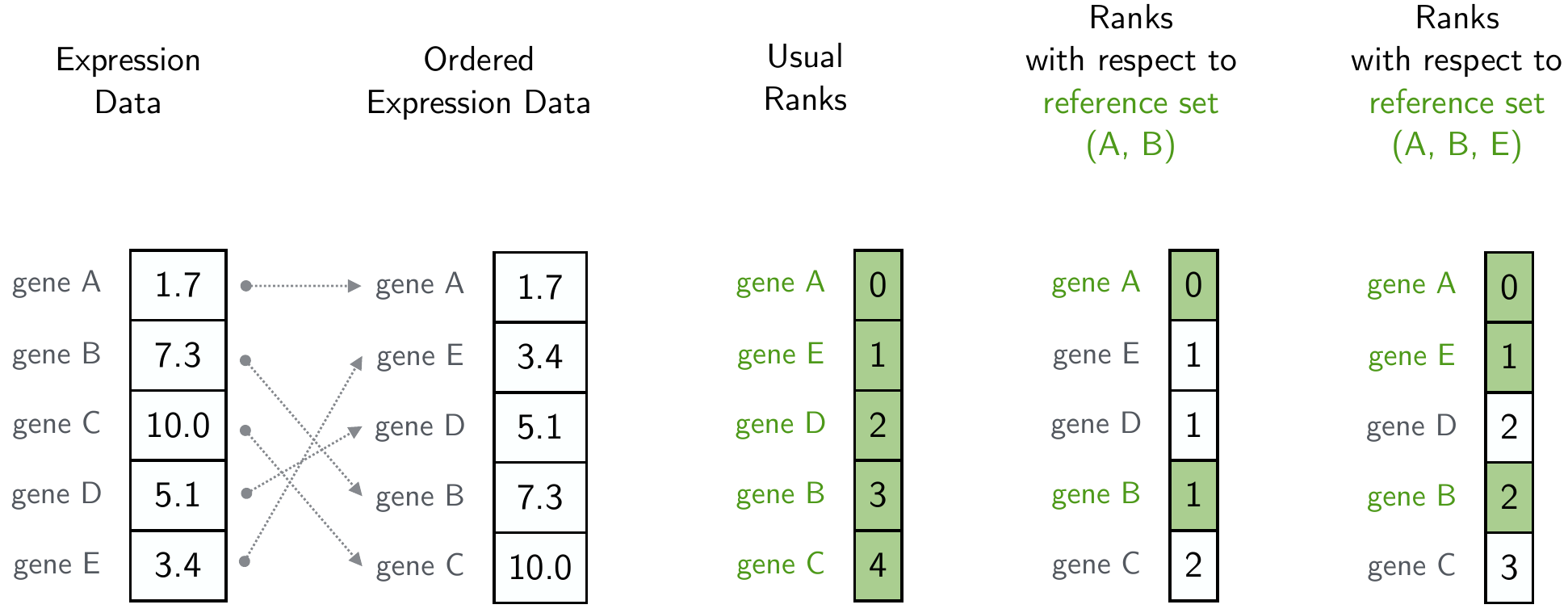} 
\caption{Example of ranking gene expression data with respect to a \emph{reference set}.}
\label{fig:example-ranking}
\end{figure}

Given a vector of gene expression levels $\x \in \RR^d,$ the rank of gene $j$, as understood in the common sense, can be obtained by counting the number of genes $k$ (among all $d$ genes) that are less expressed than gene $j$ (i.e. $x_k < x_j$).

We introduce a more general notion of rank, the rank $r_j^\Gamma$ of $x_j$ \emph{relative to a reference set} $\Gamma$, by:
\begin{align}
    r_j^\Gamma &= \sum_{k=1}^d{\1{k \in \Gamma} \1{x_j > x_k}}
\end{align}

Note that we clearly recover the classical rank in the particular case where $\Gamma=\{1,\ldots,d\}.$

In this paper, we will encode the set $\Gamma$ by a vector $\g \in \{0,1\}^d$ with $\gamma_j=\1{j \in \Gamma}.$
We can thus express $r_j^\Gamma$ as  $r_j^\Gamma=\sum_{k=1}^d{\gamma_k\, \1{x_j > x_k}}.$ To simplify notations we will also drop the exponent $\Gamma,$ which will be defined from the context.

Now, to define simultaneously all the ranks associated with the elements of a vector $\x_i$ we introduce the \emph{matrix of binary comparisons} $\C^i \in \rr^{d \times d},$ with $\C^i_{jk}=\1{x_{ij} > x_{ik}}.$ The vector of ranks $\r_i$ associated with a reference set encoded by the indicator vector $\g$ is computed as $\r_i=\C^i\g.$ 

We will use the index $i \in \{1,\dots,n\}$ to index the $n$ observations in the data.\footnote{Note that the introduced notations adopt the convention of using a strict inequality in the definition of the rank. It is possible to obtain two other similar definitions of ranks by replacing $\1{x_j > x_k}$ by $\1{x_j \geq x_k}$ or by $\1{x_j > x_k}+0.5 \: \1{x_j = x_k}.$ All definitions are obviously equivalent when there are no ties and when $\Gamma=\{1,\ldots,d\}$. The case where there are ties and where these other notions of ranks become relevant is discussed in Appendix~\ref{sec:ties-handling}.}

\subsection{Classification: learning the reference set along with a linear model on adaptive ranks} \label{sec:learning-ref-set}

We consider a classical supervised learning problem in which input data is encoded as ranks $\r_i$ of the form above, and the output data is a label $y_i \in \mathcal{Y}$. For a loss function $\ell: \mathcal{Y} \times \RR \rightarrow \RR$, we consider a regularized empirical risk minimization of the form
\begin{equation}
    \label{eq:erm}
    \min_{\w \in \rr^d,\, b \in \rr}\oon \sum_{i=1}^n \ell(y_i,\w^\top \r_i + b)+\Omega(\w),
\end{equation}
where $\Omega$ is a (typically convex) regularizer.
By expressing $\r_i$ as a function of $\g$ and of $\C^i$ and minimizing the empirical risk with respect to $\g$ as well, we propose to learn the reference set $\Gamma$.

This leads to the following optimization problem:
\begin{equation}
\label{eq:original-loss}
     \min_{\g \in \{0,1\}^d,\, \w \in \rr^{d},\, b \in \rr}\oon \sum_{i=1}^n \ell(y_i,\w^\top \C^i \g + b)+\Omega(\w).
\end{equation}

Given that the integrality constraint on $\g$ makes the optimization problem combinatorial, we propose to relax the constraint $\g \in \{0,1\}^d$ to $\g \in [0,1]^d$. We empirically found that adding a cardinality constraint on the reference set, ${\g}^{\top} \mathbf{1} = s$ instead of penalizing with a sparsity inducing regularization, such as the $\ell_1$-norm, was useful to obtain fast convergence. We suppose that this constraint removes an indeterminacy of scale between $\w$ and $\g$ which appears once the integrality constraint is removed, given that $(\alpha \, \w, \frac{1}{\alpha} \, \g)$ yields identical losses values for any scaling factor $\alpha$; it would be implicitely removed as well by regularizers on $\w$ and $\g$ but only at convergence.

Given the relaxed constraint $\g \in [0,1]^d$, and in order to nonetheless obtain solutions with $\g \in \{0,1\}^d$, we propose to solve a sequence of problems of the form
\begin{equation}
\label{eq:relaxed-loss}
     \min_{\g \in [0,1]^d,\, \w \in \rr^{d},\, b \in \rr}\oon \sum_{i=1}^n \ell(y_i,\w^\top \C^i \g + b)+\Omega(\w) + \lambda_p\, \rho(\g) \qquad \text{s.t.}\quad {\g}^{
    \top} \mathbf{1} = s,
\end{equation}
for an increasing sequence of regularization coefficients $\lambda_p,$ where $\rho$ is the concave ``push'' penalty defined by
\begin{equation}
    \rho(\g) =  \sum_{j=1}^d \gamma_j (1-\gamma_j),
\end{equation}
which effectively ``pushes'' the entries of $\g$ towards the extreme points of the hypercube $[0,1]^d.$
More precisely, starting from $\lambda_p=0$, the solution of each problem in the sequence is used as initialization to warm-start the next one, and the sequence is terminated when the solution satisfies the constraint $\g \in \{0,1\}^d$.

It is obviously possible to only solve the above problem for $\lambda_p=0$ and renounce the integrality constraints.
Actually, the presence of the capped-simplex constraints $\g \in [0,1]^d$ and $\g^\top \ones=s$ are themselves sufficient to obtain that, at the optimum, $\g^*$ tends to lie on a lower dimensional face of the capped-simplex, so that a significant fraction of its entries are exactly equal to $0$ or $1$. In preliminary experiments, we also did not observe significant differences whether integrality constraints are strictly enforced or not. The main motivations to nonetheless enforce them, are that (a) the additional computational effort is small compared to the cost of solving the problem with $\lambda_p=0,$ (b) the interpretability of the obtained ranks is otherwise lost, and (c) that it tends to produce slightly sparser solutions. 

\section{Optimization procedure} \label{sec:optimization-procedure}

\paragraph{\textbf{Block proximal coordinate descent.}} When $\lambda_p=0,$ problem~\eqref{eq:relaxed-loss} is bi-convex. More precisely, the objective function to minimize, which we denote by $\mathcal{O}(\lambda_p;\g, \w, b)$, is convex w.r.t.~$(\w,b)$ when $\g$ is fixed and convex w.r.t. $(\g, b)$ when $\w$ is fixed. This suggests that a form of alternating descent algorithm can be used, such as block coordinate descent, in which blocks of variables, here $(\g, b)$ and $(\w,b)$, are alternatively updated \citep[see for example][]{Tseng, xu2013block}. \\
In addition, since the regularizer $\Omega$ is convex and potentially non-differentiable (e.g., elastic net regularization), descent w.r.t.\ $\w$ can be suitably realized with proximal gradient steps, provided that the proximal operator for $\Omega$ can be computed efficiently.
For $\gamma$, the optimization step satisfying the constraint $\g \in [0,1]^d \mid \g^\top \ones=s$ also involves a proximal operator: the projection on this constraint set called capped-simplex. We derive this proximal operator in Appendix~\ref{sec:proximal-operator}.\\
Therefore, to solve each instance of problem~\eqref{eq:relaxed-loss}, we use a block proximal coordinate descent algorithm (BPCD). \cite{sparsebilinear} propose a BPCD algorithm to solve bilinear logistic regression problems with convex regularizers. Our implementation is similar to theirs, except that we use different blocks and a simpler stopping criterion, which is better suited to the non-convexity of the push-penalty and to the implementation of the path-following algorithm described next. We detail our implementation in Appendix \ref{app:BPCD-algorithm}.

\paragraph{\textbf{Initialization.}} Given that the optimization problem is non-convex, the initialization matters: for reasons of symmetry we set $\w=\mathbf{0}$ and $\g=\frac{s}{d} \ones,$ i.e., the center of the capped-simplex.

\paragraph{\textbf{Path-following algorithm.}} Concerning the sequence of values of $\lambda_p$ used for the problems of the form~\eqref{eq:relaxed-loss}, given that the term $\lambda_p \hspace{0.5pt}\rho$ ~eventually creates local minima at all vertices of the capped-simplex, it is important not to increase $\lambda_p$ too quickly, which could produce suboptimal solutions. We use the approach proposed by~\cite{path-following-algorithm}. In essence, we adjust the next $\lambda_p$ to ensure a sufficiently small increase of the objective value $\mathcal{O}$ for the previously found solution. The strategy is detailed in Appendix~\ref{app:path-following}.\\

To summarize, we propose to solve each instance of problem \eqref{eq:relaxed-loss} with a block proximal coordinate descent algorithm (BPCD), and to increase $\lambda_p$ according to a rule inspired by the path-following algorithm in \cite{path-following-algorithm}. This scheme is summarized in Algorithm \ref{alg:overview}.

\begin{minipage}{\textwidth}
\renewcommand*\footnoterule{}
\begin{savenotes}
\begin{algorithm}[H]
\caption{Optimization Procedure}\label{alg:overview}
\begin{algorithmic}
\State $\gamma \gets \frac{s}{d} \ones$,\; $\w \gets \mathbf{0}$, \; $\lambda_p = 0$ \Comment{Initialization}
\While{$\g \notin \{0,1\}^d$} \Comment{Iterating on $\lambda_p$}
\State $(\g, \w, b) \gets \argmin_{\g \in [0,1]^d,\, \g^\top \ones=s,\, \w \in \rr^{d},\, b \in \rr} \mathcal{O}(\lambda_p; \g, \w, b)$ \Comment{solved with BPCD}
\State $\lambda_p \gets \lambda_p^\prime$\,, with $\mathcal{O}(\lambda_p^\prime;\g, \w, b) - \mathcal{O}(\lambda_p;\g, \w, b) = \epsilon$.
\EndWhile \\
\Return $\g, \w, b$
\end{algorithmic}
\end{algorithm}
\end{savenotes}
\end{minipage}

\paragraph{\textbf{Note.}} When $\lambda_p>0$, although $\rho$ is non-convex, problem~\eqref{eq:relaxed-loss} can still be formulated as a multi-convex problem, suitable to the block coordinate descent (see Appendix~\ref{app:multi-convex formulation} for a derivation).

\subsection{Computing the product \texorpdfstring{$\C^i\g$}{C\^i(g)} with complexity \texorpdfstring{$O(d)$}{O(d)}.}

A priori, the computation of the matrix vector product $\C^i\g$ involves $d^2$ multiplications. But it is clear that to compute classical ranks it is sufficient to sort the data, which can be done with a complexity of $O(d \log d).$
Since $\C^i\g$ is none else than the vector of ranks with respect to the reference set $\Gamma,$ it seems reasonable to think that the same complexity can be achieved, and this is indeed the case.
Assuming that there are no ties, and if $\sigma_i$ is a permutation sorting the entries of $\x_i$, i.e.~ such that $x_{i,\sigma_i(1)} < \dots < x_{i,\sigma_i(d)}$,  the inner-sum can be calculated recursively by applying the same permutation to $\g$ and applying, from $j=1$ to $d$,

\begin{equation}
\label{eq:recursive-rank}
r_{i,\sigma_i(j)} \leftarrow r_{i,\sigma_i(j-1)}+\gamma_{\sigma_i(j-1)}, 
\end{equation}
with, by convention, $r_{i,\sigma_i(0)}=0.$ \\
The complexity is therefore dominated by the sorting operation and is thus $O(d \log d)$.
Moreover, sorting the data needs to be done only once at the beginning of the optimization, so that effectively the number of operation needed to compute $\C^i\g$ each time $\g$ is updated is $O(d).$ The exact same reasoning applies to the computation of $\w^\top \C^i$ which is therefore also $O(d),$ once the inverse of $\sigma_i$ is computed.
With the alternative definitions of rank and in the presence of ties, the calculations are more subtle, but the same complexity can be obtained. They are detailed in Appendix \ref{app:recursion-relation}.

\section{Benchmark: competing classification algorithms} \label{sec:classifiers}

We will apply the proposed methodology to solve a number of binary classification problems on first synthetic and then real RNA-Seq data. To serve as a basis of comparison, we choose standard logistic regression or random forest classifiers that rely either on a rank representation or not.

\paragraph{\textbf{Optirank: a sparse rank-based logistic regression with learnable reference set.}} To solve binary classification tasks, we propose \optirank, a logistic regression model on rank-transformed data, with ranks computed with respect to a learnable reference set. Our model \optirank~is fitted within the framework introduced in section \ref{sec:learning-ref-set}, by solving the optimization problem \eqref{eq:original-loss} with a logistic loss $\ell(y,a)=-y \log{(S(a))} - (1-y)\log{(1-S(a))}$, where $S(x)$ denotes the sigmoid function $S(x)=1/(1+e^{-x}),$ $y \in \mathcal{Y}=\{0,1\}$ being the binary label, and with an elastic net regularization $$\Omega(\w)=\lambda_1 \, \norm{\w}_1 + \lambda_2 \, \norm{\w}_2^2,$$ to induce sparsity in the set of features whose rank is relevant to the classification task. 
In fact, given that we consider RNA-Seq data, and that there are potentially significant correlations between genes, the use of the elastic net, with a Euclidean regularization on top of the Lasso terms, aims at stabilizing feature selection \citep[see][]{elasticnet}.

\paragraph{\textbf{Competing algorithms.}} We will compare our \optirank~ algorithm  with classical logistic regression (\lr) equipped with the same elastic net regularization $\Omega$, and logistic regression on rank-transformed data (\ranklr), still with the same regularization. In addition, in tasks involving real data, we will also compare our method to the random forest (\rf) and to the SingleCellNet algorithm (\SCN) proposed by \citet[][]{singlecellnet} for cell-typing tasks that we consider in our benchmark (see Section~\ref{sec:experiments-real-data}). The \SCN~algorithm~consists of a pre-processing pipeline which identifies gene pairs with informative differential expression and transforms the RNA-Seq data into a binary matrix indicating the order of gene pairs followed by a random forest classifier.

\paragraph{\textbf{Implementation details.}} The stopping criterion in the scikit-learn \citep{scikit-learn} implementation of logistic regression being different from the one in \optirank, to ensure that this discrepancy does not affect the comparison on real datasets, we re-optimize the weights $\w$ learned by \optirank~ with the logistic regression of scikit-learn. Additional details about the classifiers can be found in Appendix~\ref{app:classifiers}.

\section{A synthetic data distribution model with unstable ranks} \label{sec:toy-model}

In order to illustrate the potential and limitations of \optirank, we present in the following a synthetic example in which the robustness of the rank normalization is challenged. We test whether \optirank~is effectively able to overcome the difficulty of the task. 

\subsection{Model}

As mentioned in \cite{rank-based-classifiers}, the strength of rank-normalization can be linked to the fact that ranks are invariant to observation-wise monotone perturbations of the gene expressions. Those perturbations can be easily envisioned, for example by considering that counts depend in a quadratic (and observation-dependent) fashion on the RNA content in the observation. By contrast, the following example focuses on a weakness of rank normalization that arises in the presence of a non-monotone, nonetheless simple, perturbation. Those perturbations occur in real data; for instance, \cite{batch-effects-nature} report a case in which a group of genes shifts between different batches of samples. In our example, we consider a similar perturbation: we suppose that there is a group of perturbed genes, uninformative for the classification task at hand, that introduces noise in the ranks of relevant features by fluctuating in a coordinated and observation-wise manner.

More precisely, the expression levels of the genes in this group, called $P$, are all assumed to shift by an additive amount close to $\Delta_i$ (unique to the observation $i$). This introduces noise in the ranking of the other stable genes: indeed, since the perturbed genes in $P$ shift in a coordinated fashion, the rank of a stable gene is increased or decreased by an amount proportional to the number of genes in $P$ that cross it. 

We propose the following synthetic model: the non-perturbed expression of a gene $j$ in observation $i$, $\widetilde{X}_{ij}$, follows a Gaussian distribution centered on the typical expression value of gene $j$, $\mu_j$:
\begin{equation}
    \widetilde{X}_{ij} \sim \mathcal{N}(\mu_j, \sigma^2) \,,
\end{equation}
where $\sigma$ defines the magnitude of the baseline noise in the data. We generate values for $\mu_j$ by sampling uniformly on the expression interval, which we set to $\left[0,1\right]$. \\
The expression of a perturbed gene in $P$ is generated by adding to the unperturbed expression $\widetilde{X}_{ij}$ an observation-wise shift $\Delta_i$ that we sample from a centered Gaussian distribution $\mathcal{N}(0, \tau^2)$, with $\tau$ defining the typical magnitude of the perturbation. Summing all contributions, the expression of a gene $j$ in observation $i$, $X_{ij}$, is generated as:

\begin{equation}
    X_{ij} = \begin{cases}
  \widetilde{X}_{ij} + \Delta_i & \text{if } j \in P \\
  \widetilde{X}_{ij} & \text{otherwise.}
\end{cases}
\end{equation}

Finally, we assign to each observation $i$ a label generated from a simple logistic model on the ranks within the stable (non-perturbed) genes (forming the set $S$):

\begin{equation}
\label{eq:simulation-y}
    \PP(Y=1|X=\x)=\sigma(\w^\top \mathbf{r}^\Gamma + b), \quad \text{with} \quad w_j=0, \; \forall j \notin S \quad \text{and} \quad \Gamma = S.
\end{equation}

The generation of the parameters $\w$, $\g$, and $b$ is detailed in Appendix~\ref{sec:simulation-supp}.

\subsection{Results} \label{sec:toy-model-results}

We benchmarked on this synthetic data three different classifiers based on logistic regression: the simple logistic regression (\lr), the logistic regression on ranked data (\ranklr), and~\optirank, which can produce the \ranklr ~model as a particular case but offers the additional flexibility of restraining the reference set. Concerning the choice of regularization hyperparameters, note that we tuned only the ridge regularization coefficient $\lambda_2$ (and $s$ for \optirank), setting the lasso penalty $\lambda_1$ to zero for all three classifiers, given that we consider sample sizes $n$ that are large compared to the number of variables $d$. 

With default simulation parameters, \optirank~ outperforms both \ranklr~and \lr~(see Table ~\ref{tab:simulation-result}). Moreover, \optirank~ is empirically able to recover the true reference set. Indeed, the cosine similarity $S_C$ between the vector $\g$ used to generate the data (see equation \ref{eq:simulation-y}) and the one found by \optirank~ is high: $S_C = 0.95 \pm 0.03$.

\begin{table}[h!]
\centering
\begin{tabular}{|l|c|}
\toprule
    & Test Balanced Accuracy ($\%$)\\
    \hline
    Logistic Regression (\lr) & $78 \pm 2$ \\
    \hline
    Logistic Regression on Full Ranks (\ranklr) &  $80 \pm 3$\\
    \hline
    \optirank~(our model) & $96 \pm 0.4$\\
    \bottomrule
    
\end{tabular}\\[2mm]
\caption{Test balanced accuracy (in $\%$) for classifiers on the synthetic example of Section~\ref{sec:toy-model}. Default simulation parameters were set to $d=50$, $d_P=40$, $n = 1000$, $\tau = 0.2$, and $\sigma = 0.05$.}
\label{tab:simulation-result}
\end{table}

\begin{figure}[h!]
\begin{multicols}{2}
\centering
\includegraphics[width=\linewidth]{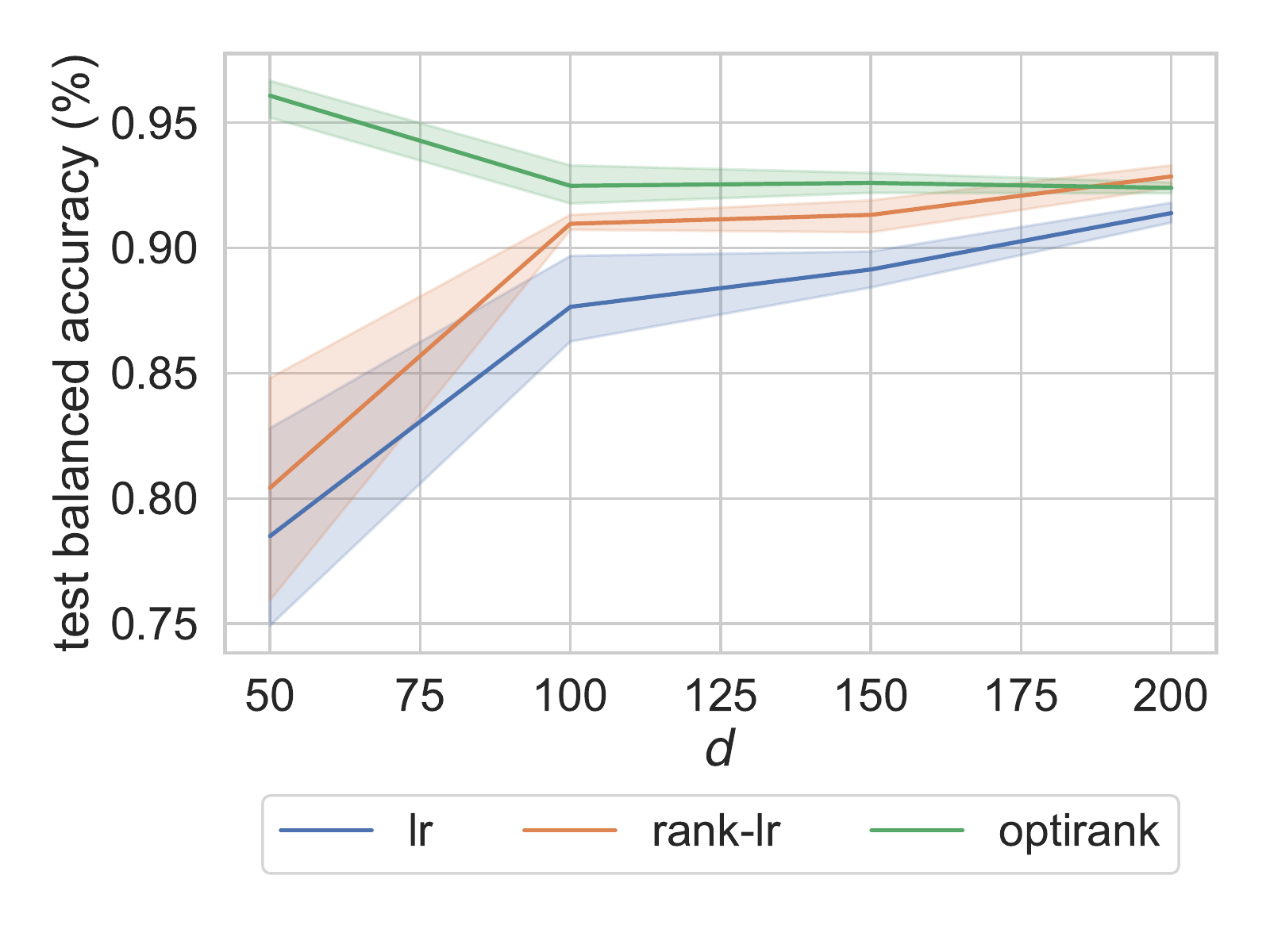}
\\
\includegraphics[width=\linewidth]{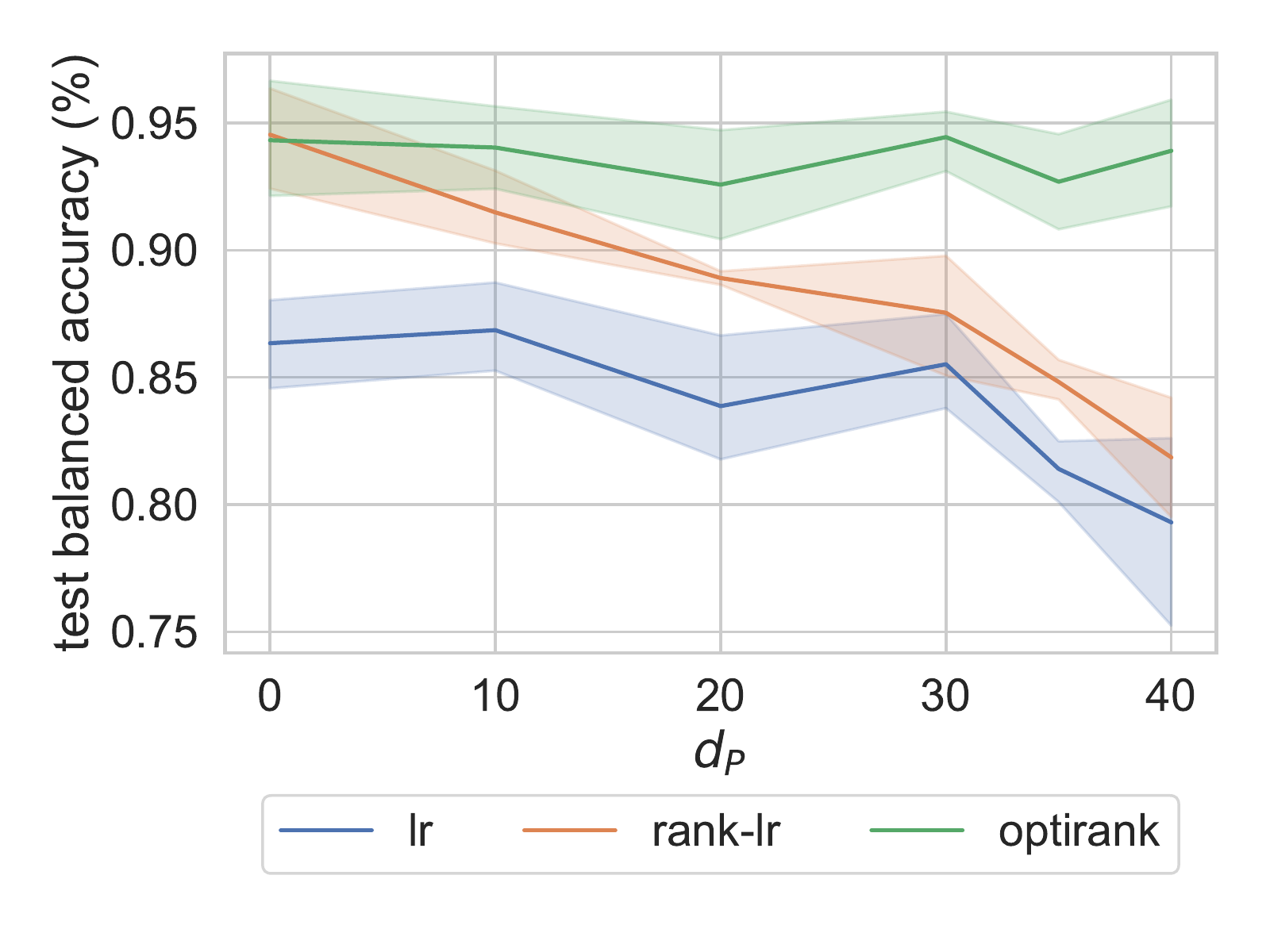}
\end{multicols}
\caption{Dependence of the test-accuracy with respect to the dimension of the gene expression profile $d$ when $d_P=40$, and with respect to the number of perturbed genes $d_P$ for $d=50$. The shaded area shows the standard error across runs.}
\label{fig:simulation-main}
\end{figure}

We investigated how this comparison evolves when we change the number of perturbed genes $d_P$ or the dimension of the gene expression profile $d$, while maintaining the ratio between the number of observations and the dimension ($n/d$) fixed. Not surprisingly, the superiority of \optirank~over \lr~and \ranklr~fades when the number of perturbed genes $d_P$ becomes small relative to the dimension of the gene expression profile. Indeed, Figure \ref{fig:simulation-main} shows that when $d$ increases while keeping the number of perturbed genes, $d_P$, equal to 40 , \ranklr~and \lr~scores rise to the level of \optirank~(whose performance degrades slightly). In accordance, when the number of perturbed genes is increased while keeping the dimension of the gene expression to 50, the performance of \ranklr~and \lr~degrades, while the score of \optirank~remains high. This outlines the fact that the perturbation on the usual ranks of informative genes becomes smaller as the ratio $d_P/d$ decreases.

Concerning the cosine similarity between the ground truth reference set and the reference set found by \optirank, its dependence on the simulation parameters $d_P$ and $d$ is small: Figure~\ref{fig:simulation-overlap} in Appendix~\ref{app:supp-figure} shows that the overlap remains high.

For the dependence of scores on $\tau$, which defines the magnitude of the observation-wise shift $\Delta_i$,  see Figure~\ref{fig:simulation-supp} in Appendix \ref{app:supp-figure}.

In summary, this synthetic task exemplifies (a) how a non-monotone perturbation can effectively degrade the performance of the rank normalization, and (b) that optirank is robust to the kind of perturbation introduced thanks to its learnable ranking reference set.

\section{Experiments on real RNA-Seq data} \label{sec:experiments-real-data}
In this section, we benchmark \optirank~on multiple biologically relevant classification tasks with real RNA-Seq data. These tasks present qualitatively different dataset shifts between train and test data. In Subsection~\ref{sec:experiment1}, we evaluate how robust different algorithms are to these dataset shifts. To enhance robustness, we then investigate in Subsection~\ref{sec:multi-source training}, an alternative learning scenario in which multiple \emph{source} datasets are merged in the training data. In this manner, we hope that the algorithm learns better to be robust to the kind of perturbation it will encounter in the test data.

\subsection{Classification with different dataset shifts} \label{sec:experiment1}

\subsubsection{Classification tasks} \label{sec:classification-tasks}

\paragraph*{\textbf{CUP-related tasks.}} Cancer of unknown primary (CUP) occurs when a patient has a metastatic tumor whose organ of origin (where the primary tumor was located) cannot be determined. A lot of effort has been dedicated to develop classifiers to predict the organ of the primary tumor based on RNA-Seq data of the metastatic tissue, in the hope of personalizing and enhancing the treatment given to CUP patients \citep{CUP-review}. However, an obstacle in building efficiently such classifiers is the scarsity of RNA-Seq data of metastatic tumors. As a result, classifiers are often trained and tuned on datasets of primary tumors, which are biologically different from metastasis, and metastatic samples are reserved to the external classifier validation. This is precisely what we do in the three tasks \task{TCGA}, \task{PCAWG}, \task{met500}. In the task \task{TCGA}, classifiers are trained on the TCGA dataset comprising primary tumors \citep{TCGA}, and tested on a held out portion of the same dataset. In the task \task{PCAWG}, those same classifiers trained on TCGA were tested on an external dataset PCAWG also with primary tumors \citep{PCAWG}. Finally, the task \task{met500} evaluates those classifiers on the met500 dataset \citep[published by][]{met-500}, which comprises metastatic tumors of various origins. The two latter tasks represent different challenges in terms of dataset-shift between train and test. The task \task{PCAWG} is subject to technical variation between two separately obtained datasets, which we call \emph{batch-effect}. An additional difficulty in the \task{met500} task is that a metastasis differs biologically from its primary tumor, resulting in a so-called \emph{biologically induced dataset shift}.

\paragraph*{\textbf{Cell-typing single-cell data.}} Single-cell RNA-Seq (scRNA-Seq) provides a way to probe gene expression at the cell-level resolution. A preliminary step in single cell data analysis resides in the cell-type identification of each observation/cell, i.e. cell typing. To achieve this automatically, \cite{singlecellnet} propose to train a classifier on a labeled dataset comprising many cell types ---\hspace{1pt}commonly known as a cell-atlas, and use it to infer the cell-types in an unlabelled dataset. One difficulty is that the train and test scRNA-Seq data are potentially generated by different sequencing platforms. \cite{singlecellnet} evaluate the robustness of their classifier \SCN~for various cross-platform train and test data combinations (Tasks \task{Baron-Murano}, \task{Baron-Segerstolpe}, \task{MWS-TM10x}, \task{MWS-TMfacs}, \task{TM10x-MWS}, \task{TM10x-TMfacs} and \task{TMfacs-MWS}). We use the same tasks to investigate the usefulness of \optirank~in counteracting \emph{dataset shifts induced by different sequencing platforms} and compare with the original method \SCN. 
\paragraph*{\textbf{\task{BRCA} task.}} The task \task{BRCA} consists in predicting the presence of the BRCA mutation from the RNA-Seq data in breast primary tumors from the TCGA dataset. This task does not directly answer a real-life classification problem. However, the classifier coefficients could be used to obtain a (sparse) transcriptional signature for BRCA cancer.

We list all tasks while grouping them by type of  dataset shift (if any) between the train and test data in Table~\ref{tab:tasks-dataset-shift}.

\begin{table}[h!]
\centering
\begin{tabular}{|p{5cm}|p{8cm}|}
\toprule
   \textbf{Dataset-shift} & \textbf{Tasks}\\
    \midrule
    None (same distribution) & \task{BRCA}, \task{TCGA} \\
    \hline
    Batch-effects &  \task{PCAWG}\\
    \hline
    Technical dataset-shift (Different sequencing platforms) & \task{Baron-Murano}, \task{Baron-Segerstolpe}, \task{MWS-TM10x}, \task{MWS-TMfacs}, \task{TM10x-MWS}, \task{TM10x-TMfacs} and \task{TMfacs-MWS} \\
    \hline Biologically induced dataset-shift & \task{met500} \\
    \bottomrule
    
\end{tabular}\\[2mm]
\caption{Classification tasks by type of dataset-shift between train and test set.}
\label{tab:tasks-dataset-shift}
\end{table}

Additional details about the data sources are in Appendix \ref{app:datasets}.

\subsubsection{Data pre-processing}

For fairness of comparison and as first step of dimensionality reduction, the data was reduced to include the 1000 genes occuring in informative pairs identified by \SCN. Logged raw-cpm values were used as input for each classifier (see Appendix \ref{app:datasets-preprocessing} for additional details).

\subsubsection{Hyperparameter selection}

The $\ell_1$ and $\ell_2$ regularization coefficients of \optirank, \lr ~ and ~\ranklr ~, and the $s/d$ ratio for \optirank~ were tuned via internal cross-validation (i.e., using held out data from the same source as the training data; see Appendix \ref{app:datasets-cv1}, \ref{app:CV-multi-source} and \ref{app:results-hp-grid} for details). SingleCellNet was used with parameters suggested by \cite{singlecellnet}, and random forest (\rf)~was trained with $300$ trees. Optimal hyperparameters were chosen with the one-standard error rule \citep{one-std} which selects the sparsest model with a score within one standard error of the best one. 

\subsubsection{Results and discussion} \label{sec: results-and-discussion}

The performance in terms of balanced accuracy of all classifiers on the different classification tasks presented in Section~\ref{sec:classification-tasks} are summarized in the following table (Table~\ref{tab:summary-results-simple_tasks}). We highlighted in bold the scores of classifiers that did not score significantly worse than the winning method, according to a paired Student's t-test.

\begingroup
\setlength{\tabcolsep}{3pt}
\begin{table}[h!]
\centering
\addtolength{\leftskip} {-2.5cm} 
\addtolength{\rightskip}{-2.5cm}
\label{tab:summary-results-simple_tasks}
\begin{tabular}{|l|ccccc|}
\toprule
{} &                                  \SCN &                                     \lr &                               \optirank &                               \ranklr &                   \rf \\
\midrule
\task{BRCA}              &                    $50.1 \pm 0.3$ (4) &        $\mathbf{58 \pm 2}$ \textbf{(1)} &                          $52 \pm 2$ (3) &      $\mathbf{53 \pm 1}$ \textbf{(2)} &      $50.0 \pm 0$ (5) \\
\task{TCGA}              &                        $92 \pm 1$ (5) &  $\mathbf{99.14 \pm 0.23}$ \textbf{(2)} &  $\mathbf{99.06 \pm 0.26}$ \textbf{(3)} &  $\mathbf{99.3 \pm 0.3}$ \textbf{(1)} &    $95.7 \pm 0.5$ (4) \\
\hline
\task{PCAWG}             &  $\mathbf{76.0 \pm 4.5}$ \textbf{(2)} &    $\mathbf{76.2 \pm 5.4}$ \textbf{(1)} &    $\mathbf{74.3 \pm 5.4}$ \textbf{(4)} &  $\mathbf{74.4 \pm 5.8}$ \textbf{(3)} &        $60 \pm 4$ (5) \\
\hline
\task{met-500}           &                        $67 \pm 4$ (4) &                          $71 \pm 4$ (3) &        $\mathbf{77 \pm 4}$ \textbf{(1)} &      $\mathbf{75 \pm 4}$ \textbf{(2)} &        $60 \pm 4$ (5) \\
\hline
\task{Baron-Murano}      &                        $89 \pm 2$ (3) &                          $87 \pm 3$ (4) &    $\mathbf{93.0 \pm 1.9}$ \textbf{(2)} &  $\mathbf{93.2 \pm 1.9}$ \textbf{(1)} &        $62 \pm 3$ (5) \\
\task{Baron-Segerstolpe} &  $\mathbf{93.5 \pm 1.6}$ \textbf{(3)} &    $\mathbf{93.4 \pm 2.2}$ \textbf{(4)} &    $\mathbf{93.6 \pm 2.2}$ \textbf{(2)} &  $\mathbf{94.0 \pm 1.8}$ \textbf{(1)} &        $60 \pm 3$ (5) \\
\task{MWS-TM10x}         &                        $72 \pm 2$ (4) &        $\mathbf{86 \pm 2}$ \textbf{(1)} &        $\mathbf{84 \pm 2}$ \textbf{(3)} &      $\mathbf{85 \pm 2}$ \textbf{(2)} &        $53 \pm 1$ (5) \\
\task{MWS-TMfacs}        &                        $70 \pm 2$ (4) &        $\mathbf{86 \pm 1}$ \textbf{(3)} &    $\mathbf{87.3 \pm 1.6}$ \textbf{(2)} &  $\mathbf{87.4 \pm 1.7}$ \textbf{(1)} &  $50.01 \pm 0.01$ (5) \\
\task{TM10x-MWS}         &                    $51.5 \pm 0.3$ (4) &        $\mathbf{72 \pm 2}$ \textbf{(1)} &                          $65 \pm 2$ (2) &                        $63 \pm 2$ (3) &    $51.0 \pm 0.2$ (5) \\
\task{TM10x-TMfacs}      &                        $80 \pm 1$ (4) &                          $91 \pm 1$ (3) &    $\mathbf{92.3 \pm 0.9}$ \textbf{(2)} &  $\mathbf{92.7 \pm 0.9}$ \textbf{(1)} &        $58 \pm 1$ (5) \\
\task{TMfacs-MWS}        &                    $51.0 \pm 0.2$ (4) &        $\mathbf{71 \pm 2}$ \textbf{(1)} &                          $64 \pm 1$ (3) &                        $66 \pm 2$ (2) &    $50.2 \pm 0.1$ (5) \\
\bottomrule
\end{tabular}\\[2mm]
\caption{Average balanced accuracies in \% (across folds and classes) of competing classifiers on the different tasks detailed in Section~\ref{sec:classifiers}. Horizontal lines separate tasks with different types of dataset shift, from top to bottom: generalization to the same distribution, robustness to batch-effects, robustness to biologically-induced dataset shifts and robustness across sequencing platforms. The integer in parenthesis denotes the rank of the classifiers in terms of average balanced accuracy (lower is better). Classifiers which did not score significantly worse than the best classifier according to a paired Student's t-test (with a level of 5 \%) are highlighted in bold (see Appendix \ref{app:comparison-between-classifiers} for additional details).}
\end{table}

\endgroup

Interestingly, the advantage of logistic regression-based classifiers relying on a rank-representation (\optirank, \ranklr) over their non-ranked counterpart (\lr) is not consistent, but rather depends on the task considered. Indeed, on the tasks \task{TM10x-MWS} and \task{TMfacs-MWS}, \lr~clearly surpasses its ranked counterparts, while on the tasks \task{met500}, \task{Baron-Murano} and \task{TM10x-TMfacs} we notice the opposite trend. This indicates that the rank representation confers additional robustness against dataset shifts only in some instances.

A burning question is whether there is an advantage of ranking relative to a subset of genes compared to ranking among all. At first sight, this doesn't seem to be the case: the performances of \optirank~and \ranklr~are similar. In a more thorough analysis in which we carried paired Student's t-tests for every task and every pair of classifiers (see Appendix \ref{app:comparison-between-classifiers}), only the task \task{TM10x-MWS} showed a significant difference between \ranklr~and \optirank, in favor of \optirank. 

In summary, in these tasks, the ranking reference set found by \optirank~is not more robust than the classical full reference set. A possible explanation is that an optimal restricted reference set does not necessarily exist. Contrarily to the synthetic example in Section \ref{sec:toy-model} and to certain observations made on real data \citep[see][]{batch-effects-nature}, where a group of genes shift in one direction and perturb the ranking, in the tasks we consider, the dataset-shift could either be a monotone transformation or could shift genes in opposite directions. In both these scenarios, the ranks of certain stable genes would not be affected by the dataset shift. Alternatively, one could argue that even if such an optimal restricted reference set existed, the only way to discover it would be by inspecting the test dataset. We address this question in an additional experiment presented in the next section. 

Aside from the performance aspect, it is important to note that by definition, the classical ranking normalization is computed with the measurement of all (reference) genes. In contrast, \optirank~can find models that require only a small number of genes to be sequenced, which can be a decisive advantage in some medical applications. In the tasks we consider, solutions found by \optirank~require around 500 genes, half of the thousand used by \ranklr. However, it is worth noting that when the logistic regression performs well, there is no advantage of using \optirank, as the latter tends to produce less sparse solutions (see Appendix \ref{sec:sparsity-investigation}).

It is worth noting that in general, the random forest \rf~performs worse than other classifiers, and that \SCN~does not provide a competitive advantage on single-cell typing tasks.

\subsection{Enhancing robustness with a multi-source learning scenario}\label{sec:multi-source training}

In this experiment, we investigate whether in the presence of dataset shifts, merging two \emph{source} datasets in the training set increases the classification accuracy on the third external \emph{target dataset}. The rationale is that the algorithm could learn to be robust to the kind of perturbation it will encounter in the \emph{target dataset}\footnote{The art of combining multiple labeled source datasets in order to classify a target dataset under a dataset-shift is referred as \emph{multi-source domain adaptation} in the literature (See for example the review by \cite{multi-source}).}. \\
To achieve this, we constructed three tasks: \task{TCGA-PCAWG-met500}, \task{Baron-Segerstolpe-Murano} and \task{MWS-TMfacs-TM10x}. The tasks are named after the datasets that compose them: the first is the main \emph{source} dataset, the second the \emph{auxiliary source} dataset and the third is the \emph{target} dataset. We compare the performance in the \emph{multi-source} scenario (where the \emph{main} and \emph{auxiliary} source datasets are merged into a training set) to a baseline scenario in which the \emph{auxiliary source} dataset is not used (\emph{single-source} scenario). Appendix \ref{app:CV-multi-source} provides additional details about the construction of those tasks.

\paragraph{\textbf{ANrank-lr.}} One could ask if, with the help of the \emph{auxiliary source} dataset, robust ranking reference genes can be identified in a simpler manner than in \optirank, in particular, with a selection step decoupled from the fitting process. To answer this question, we constructed an additional logistic regression classifier based on adaptive ranks, \ANranklr, that selects the ranking reference genes based on a simple ANOVA test. For each gene, the \emph{vulnerability} to dataset-shift is assessed with a two-way ANOVA which determines the effect of label and dataset jointly. The $s$ most robust genes are selected as ranking reference (See Appendix \ref{app:ANranklr} for additional details). For completeness, we evaluate \ANranklr~both in the \emph{multi-source} and in the \emph{single-source} scenario. In the \emph{single-source} setting, \ANranklr~uses the \emph{auxiliary source} dataset only for the ANOVA test (and not during fitting nor validation).

\paragraph{\textbf{Data preprocessing and hyperparameter selection.}}

Data preprocessing was done as described in the previous section. Appendix \ref{app:CV-multi-source} details the procedure to obtain the cross-validation splits: special care was taken to have similar training dataset sizes in the \emph{multi-source} and \emph{single-source} scenarios. The hyperparameter grids used for cross-validation are the same as in the previous experiment. Concerning \ANranklr, the number of ranking reference genes $s$ and the elastic net regularization coefficients are tuned over the same grid as for \optirank.

\subsubsection{\textbf{Results and Discussion.}}
There is a clear benefit brought by merging two source datasets in the training phase (\emph{multi-source} scenario). Indeed, for nearly all classifiers and tasks, the average balanced accuracy is greatly increased in the \emph{multi-source} scenario (Table \ref{tab:summary-results-multiple_sources}) compared to the \emph{single-source} scenario in which the \emph{auxiliary source} dataset is not used (see Table \ref{tab:summary-results-single_source} in Appendix \ref{app:results-single-source}). Accordingly, for both single-cell typing tasks, the leading score is greater in the \emph{multi-source} scenario than in the \emph{single-source} scenario. Moreover, the regression classifiers based on ranks (\optirank, \ANranklr, and \ranklr) outperform the simple logistic regression (\lr).

However, as in the previous section, the performances of \optirank~and \ranklr~seem comparable: in the single-cell tasks, the paired t-tests do not reveal any significant difference between the two classifiers (see Appendix \ref{app:comparison-between-classifiers}).

It is worth noting that despite the simplicity of its method for restricting the reference set, \ANranklr~reaches a level of performance comparable with the other ranked-based algorithms, in particular \optirank, and likewise outperforms the simple logistic regression. This is particularly interesting since, by definition, \ANranklr~can produce sparse solutions. Indeed, in Appendix \ref{sec:sparsity-investigation-multi-source}, we note that \optirank~and \ANranklr~find solutions involving a similar number of genes. In accordance with the results of the previous section, the simple logistic regression produces substantially sparser solutions.

\begingroup
\setlength{\tabcolsep}{2pt}
\begin{table}[h!]
\centering
\addtolength{\leftskip} {-3cm} 
\addtolength{\rightskip}{-3cm}
\resizebox{\columnwidth}{!}{%
\label{tab:summary-results-multiple_sources}
\begin{tabular}{|l|cccccc|}
\toprule
{} &                               \ANranklr &                \SCN &                               \lr &                               \optirank &                               \ranklr &             \rf \\
\midrule
\task{TCGA-PCAWG-met500}        &        $\mathbf{81 \pm 4}$ \textbf{(1)} &      $69 \pm 5$ (4) &  $\mathbf{80 \pm 4}$ \textbf{(2)} &                          $68 \pm 4$ (5) &                        $71 \pm 4$ (3) &  $65 \pm 4$ (6) \\
\task{Baron-Segerstolpe-Murano} &                      $98.0 \pm 0.3$ (3) &  $93.2 \pm 0.8$ (5) &                $97.2 \pm 0.4$ (4) &    $\mathbf{98.1 \pm 0.3}$ \textbf{(2)} &  $\mathbf{98.2 \pm 0.3}$ \textbf{(1)} &  $67 \pm 3$ (6) \\
\task{MWS-TMfacs-TM10x}         &  $\mathbf{95.63 \pm 0.40}$ \textbf{(3)} &      $83 \pm 1$ (5) &                $92.5 \pm 0.9$ (4) &  $\mathbf{95.64 \pm 0.41}$ \textbf{(2)} &  $\mathbf{95.9 \pm 0.4}$ \textbf{(1)} &  $72 \pm 2$ (6) \\
\bottomrule
\end{tabular}}\\[2mm]
\caption{\textbf{\emph{Multi-source scenario}.} Average balanced accuracies in \% (across folds and classes) of competing classifiers on the tasks detailed in section~\ref{sec:multi-source training} in the case in which the first two \emph{source datasets} are merged in the training phase. The integer in parenthesis denotes the rank of the classifiers in terms of average balanced accuracy (lower is better). Classifiers which did not score significantly worse than the best classifier according to a paired Student's t-test are highlighted in bold (see App. \ref{app:comparison-between-classifiers} for additional details).}
\end{table}

\endgroup

\paragraph*{\textbf{Runtime comparison.}} The runtime for competing classifiers was measured in the previous \emph{single-source} scenario. Table \ref{tab:runtime-estimates} in Appendix \ref{app:runtime-estimates} attests that the fitting time of both \optirank~and \ANranklr~are reasonable and in some instances lower than the one of their competitors.

\section*{Conclusion}

According to the literature, rank normalization confers increased robustness against distribution shifts that occur in RNA-Seq data. This success is linked to the fact that rank normalization is invariant to all perturbing monotone transformations that occur between different datasets and/or samples. However, a potential weakness of using rank normalization is that the rank of genes that might be biologically relevant can be perturbed by fluctuations of irrelevant ones. \\
To counteract this problem, we proposed \optirank, an algorithm that learns a ranking relative to an optimal reference genes set while learning a classification or regression model. We showed on a synthetic example, inspired by observations on real data, how rank-normalization can suffer from collective fluctuations of an ensemble of genes that perturb the ranks, and demonstrated the ability of \optirank~to eliminate those genes from the ranking reference set, thereby allowing it to solve successfully the classification task.

We then assessed the performance of \optirank~on 11 real classification tasks, presenting different challenges in terms of distribution shifts occurring between train and test data. Indeed, we hypothesize that our model is able in some instances to remove from the ranking reference set genes that have the propensity to shift in the test distribution, thereby perturbing the ranks learned on the training data. Firstly, we observed that the advantage of the rank transformation is not systematic. Moreover, contrary to our hypothesis, restricting the reference set, as is done by \optirank, does not seem to provide increased robustness compared to ranking relative to the full set of genes.

As an additional way to tackle distribution shifts occurring between train and test data, we propose a \emph{multi-source} learning scheme. In this scheme, we train a classifier on a union of two different datasets in which a dataset shift occurs, hoping to make it more robust and efficient on a third external dataset. We show that this scenario is particularly useful in the cell-typing tasks, in particular when used in synergy with rank-based classifiers. We also explored an alternative way of restricting the reference set, with a simple ANOVA test that exploits the multiple sources in the training data. Despite its simplicity, the resulting classifier, which we call \ANranklr, achieves a level of performance similar to \optirank.

Finally, it is important to mention that restricting the reference set reduces the number of genes needed to be sequenced, while maintaining the level of robustness and accuracy of the rank normalization. Therefore, in certain medical applications where sparsity is desired, it can be worth considering the classifiers \optirank~and \ANranklr.
\medskip

\section*{Acknowledgements}
This work was funded under the Swiss Data Science Center collaborative project grant C19-02.

\bibliographystyle{apalike}
\bibliography{main}

\section{Code and Data Availability} \label{sec:code-availability}

The code and the data necessary to reproduce the results are available on the Github repository \url{https://github.com/paolamalsot/optirank}.


\clearpage

\appendix

\section{Notions of rank in the presence of ties} \label{sec:ties-handling}
For the RNA-Seq data we consider in several experiments, the read count of a few genes are equal to $0$. This leads to ties between these genes, which motivated us to extend the formulation proposed to that case.  

\subsection{Different rank definitions}
Ever since ranks were introduced in statistics, there have been discussions on how to correctly treat ties~\citep{kendall1945treatment}. For more references and discussion on recent related work, we refer the reader to \citet{amerise2015correction}.

The three simplest approaches, and the ones which could be relevant in our setting, consist in assigning to a group of tied genes whose ties are initially broken arbitrarily, respectively their \emph{minimum}, \emph{maximum}, or \emph{average} rank.

For simplicity, and given that the problem of ties is not central to the set of ideas that we are presenting and might be irrelevant in many cases, the generalization of classical ranks to ranks \emph{with respect to reference set} that we introduce in the paper stems from the \emph{minimum rank} definition. However, our implementation of the \optirank~algorithm uses the better behaved \emph{average} rank.

We propose in the following generalizations of these different ranks to the case where ranks are defined with respect to a reference set $\Gamma.$

\paragraph{Minimum rank.} In Section~\ref{sec:ranking-ref-main} we introduced the following definition of rank $r_j^\Gamma$ of $x_j$ \emph{relative to a reference set} $\Gamma$:
\begin{align}
    r_j^\Gamma &= \sum_{k=1}^d{\1{k \in \Gamma} \1{x_j > x_k}} \;.
\label{eq:min-ranking}  
\end{align}
The above definition of ranks assigns to tied values the minimum rank they would have if they were arbitrarily ordered, hence the name of \emph{minimum rank}. This rank is also referred to as the \emph{standard competition rank} for obvious reasons. Note that with the mathematical definition above, the smallest rank equals $0$; to obtain ranks that starts at $1$ it suffices to add $1$ to all rank values. 

\paragraph{Average rank.} A way of handling ties which has better properties, in particular which keeps constant the sum of the ranks, is to assign to them the average value of these ranks, hence the name \emph{average ranking}. In mathematical terms, this is:

\begin{align}
    r_i^{\Gamma} &= \sum_{j=1}^d{\1{j \in \Gamma} \left( \1{x_i > x_j} + 0.5\,  \1{x_i = x_j}\right)} - 0.5\\
          &= \sum_{j=1}^d{\gamma_j \left( \1{x_i > x_j} + 0.5\,  \1{x_i = x_j}\right)} - 0.5 \,,
\label{eq:avg-ranking}  
\end{align}
where the offset of $0.5$ sets again the lowest rank to the value of $0$. (The fact that this offset appears here while it did not appear for the minimum rank could seem surprising, but it is necessary because we sum over all values of $j$ including $j=i$.) Again to obtain ranks that starts at $1$ we can add $1$ to all ranks. The \emph{average rank} is also sometimes called the \emph{fractional rank}.

\paragraph{Maximum rank.} Yet another possible definition, which is symmetric with the minimum rank, consists in replacing the strict inequality in \eqref{eq:min-ranking} by an inclusive inequality and adding an offset of $1$ (to keep our convention that the ranks start at $0$). This results in:

\begin{align}
    r_j^\Gamma &= \sum_{k=1}^d{\1{k \in \Gamma} \1{x_j \geq x_k}} - 1 \;.
\label{eq:max-ranking}  
\end{align}
The \emph{maximum rank} is also called \emph{modified competition rank}.

In our implementations and experiments, we systematically used the \emph{average rank} for two main reasons. First, if two variables (i.e.~genes in our case) have systematically close values (i.e.~that are either equal or tend to be very close and in arbitrary order), we would expect the coefficients $w_j$ associated with their ranks in a classification model to be close. In that case, it is natural to request that the linear score does not change much whether they are exactly equal or not. Given that the sum of the ranks of tied values is constant for the average rank, it satisfies this property, which fails for the min and the max rank. Second, for a non-trivial reference set, and when there are no numerical ties, the average rank only assigns integer valued ranks for the elements of the reference set $\Gamma$, and any element falling exactly in-between two consecutive reference elements has a half-integer value, which is a nice property to have.

\subsection{Recursion to compute \emph{average/maximum/minimum ranks} in \texorpdfstring{$O(d)$}{O(d)} from sorted data} \label{app:recursion-relation}

In this section, we derive the recursion used to compute the rank with respect to a reference gene set in the presence of ties for all definitions of ranking mentioned previously.

Let $\x \in \rr^d$ represent the vector of gene expressions and $\Pi = (A_1, \dots, A_K)$ its ordered partition, such that
\begin{equation}
\begin{dcases}
\begin{aligned}
&\forall k \in \{1,\dots,K\}, \, \forall i, i' \in A_k, \; & x_i = x_{i'}, \\
&\forall j<k, \, \forall i\in A_j, i' \in A_k, \; & x_i < x_{i'}.
\end{aligned}
\end{dcases}
\end{equation}

\subsubsection{Recursion for the average rank}

The recursive relationship for the rank of gene $i \in A_{k_i}$ is derived as follows, from the definition in Eq.~\eqref{eq:avg-ranking}:

\begin{align}
    r_i &= \sum_{j=1}^d{\gamma_j \left( \1{x_i > x_j} + 0.5\,  \1{x_i = x_j}\right)} - 0.5 = \sum_{k=1}^K \sum_{j \in A_k} \gamma_j \big(\1{k_i > k} + 0.5\, \1{k_i = k}\big) - 0.5 \\
    &=\sum_{k=1}^{k_i} \abs{\Gamma \cap A_k} \big(\1{k_i > k} + 0.5\, \1{k_i = k}\big) - 0.5,
\end{align}
where $k_i$ denotes the index of the partition such that $i \in A_{k_i}.$

Obviously, $r_i=r_{i'}$ if $k_{i}=k_{i'}.$
If we call $\widetilde{r}_{k_i}$ this common value, with a little work, we obtain the following recursive relationship:

\begin{equation}
    \widetilde{r}_{k} = \widetilde{r}_{k-1} + \cfrac{1}{2} \big(\abs{\Gamma \cap A_{k -1}} + \abs{\Gamma \cap A_{k}} \big) + 0.5.
\end{equation}

\subsubsection{Recursion for the maximum rank}

The recursion for the \emph{maximum} rank strategy defined by Eq.~\eqref{eq:max-ranking} can be obtained similarly. With the same notations, we have  

\begin{equation}
    \widetilde{r}_k = \widetilde{r}_{k-1} + \abs{\Gamma \cap A_{k}}.
\end{equation}

\subsubsection{Recursion for minimum rank}

The recursive relation for the minimum rank accounting for the presence of ties can 
be obtained similarly. With the same notations, we have:

\begin{equation}
    \widetilde{r}_k = \widetilde{r}_{k-1} + \abs{\Gamma \cap A_{k-1}}.
\end{equation}


\section{Optimization algorithms}
\subsection{Projection on the capped-simplex} \label{sec:proximal-operator}
To be self-contained, we rederive in this section an efficient algorithm to compute the projection on the capped-simplex. Note that it is a classical result that this projection can be calculated in $O(d log d)$ operations, as demonstrated in \cite{capped-simplex}. 

Suppose that for a given $\x$ in $\rr^d$, we wish to solve:

\begin{equation}
\label{eq:objective-convex}
    \min_{\z \in \rr^d} \cfrac{1}{2} \norm{\z - \x}^2_2, \quad \text{such that } \quad 0 \le z_i \le 1 \quad \text{and} \quad \sum_{i=1}^d {z_i} = k.
\end{equation}

Because of the convexity of Problem~\eqref{eq:objective-convex}, any point satisfying the KKT conditions is primal optimal (and vice versa).

The Lagrangian is:
\begin{equation}
    \mathscr{L}(\z, \lambda, {\boldsymbol \nu}, {\boldsymbol \mu}) = \cfrac{1}{2} \norm{\z - \x}^2_2 + \lambda \big(\sum_{i=1}^d {z_i} - k\big) - \sum_{i=1}^{d} {\nu_i z_i} + \sum_{i=1}^{d}{\mu_i(z_i - 1)}.
\end{equation}

We enforce the KKT conditions, namely complementary slackness (CS), primal feasibility (PF), dual feasibility (DF), and primal stationarity (PS), to get

    \begin{align}
     \forall i \in \{1,\ldots d\}, \quad 
    & \begin{dcases}
    \begin{aligned}
    &\nu_i z_i = 0, \\
    &\mu_i (z_i -1) = 0,
    \end{aligned}
    \end{dcases}\tag{CS}\\
    & \sum_{i=0}^d{z_i} = k, \tag{PF}\\
     \forall i \in \{1,\ldots d\}, \quad  & \nu_i, \mu_i  > 0, \tag{DF}\\
    & \nabla_{\z} \mathscr{L} = 0.
    \tag{PS}
    \end{align}

With a little work, we get that:
\begin{equation}
    z_i = \clip(x_i - \lambda, 0, 1),
\end{equation} 
with $\clip(x, a, b) = \max(a, \min(x, b))$ (provided $a<b$) and where $\lambda$ is found by solving $\psi(\lambda)=k$, with:
\begin{equation}
    \label{eq:psi-equation}
    \psi(\lambda) = \sum_{i=0}^d \clip(x_i - \lambda, 0, 1).
\end{equation}

In order to find the solution of \eqref{eq:psi-equation}, one can follow this procedure:

\begin{enumerate}
\item Order $\x_i$ and $\x_i-1$ values (yielding $2d$ ordered values or knots). As a result, there are $2d-1$ \emph{intervals} between subsequent knots.
\item Calculate the slope of $\psi(\lambda)$ on each \emph{interval}.
\item Calculate iteratively the value of $\psi(\lambda)$ at each knot, starting from the greatest $\x_i$ where $\psi(\lambda)=0$. One can stop when $\psi(\lambda)>k$.
\item Determine the \emph{interval} on which $\psi(\lambda)=k$.
\item Solve the linear equation on this \emph{interval} to find $\lambda$.
\end{enumerate}

\subsection{Adaptive step sizes and a multi-convex formulation} \label{app:multi-convex formulation}

In this section, we detail the exact problem formulation to which a block proximal coordinate descent algorithm~\citep[BPCD,][]{xu2013block} can be applied. Our formulation and algorithm is in particular very close to the ones considered in~\citet{sparsebilinear}.

As stated in Section~\ref{sec:optimization-procedure}, the goal is to find the solution of a minimization problem of the form:
\begin{equation}
    \label{eq:relax-2}
     \min_{\g \in [0,1]^d,\, \w \in \rr^{d},\, b \in \rr}\oon \sum_{i=1}^n \ell(y_i,\w^\top \C^i \g + b)+\Omega(\w) + \lambda_p\, \rho(\g) \qquad \text{s.t.}\quad {\g}^{
    \top} \mathbf{1} = s,
\end{equation}
for an increasing sequence of coefficients $\lambda_p \ge 0$, where $a\mapsto \ell(y,a)$ is assumed to be a convex and differentiable loss function, with Lipschitz gradients, where $\Omega(\w)$ is a convex regularizer, and where $\rho$ is the concave ``push'' penalty defined by
\begin{equation}
    \rho(\g) =  \sum_{j=1}^d \gamma_j (1-\gamma_j).
\end{equation}
When $\lambda_p>0$, the previous problem is clearly non convex w.r.t.~$\g$. It is obviously possible to use block-coordinate proximal coordinate descent on functions that are not even convex with respects to each of the individual blocks considered~\citep{razaviyayn2013unified,csiba2017global}. But, with among others the motivation of being able to use adaptive step-sizes that can be defined in a principled way, we propose to exploit the fact that the optimization problem can be reformulated as another bi-convex problem, by introducing a variational form for the concave penalty $\rho$.

Indeed, thanks to the Fenchel-Legendre transform it is possible to express the concave push-penalty as an infimum over a set of linear functions parameterized by the dual variable $\Bu$, as follows
\begin{align}
    &\rho(\g) = \inf_{\Bu \in \rr^d} \big\{ \rho^*(\Bu) - \g^\top \Bu\big\},  \quad \text{with} \quad
     \rho^*(\Bu) = \sum_{j=1}^d{\textstyle{\frac{1}{4}} (1 + \upsilon_j)^2}.
\end{align}
Therefore, the minimization problem \eqref{eq:relax-2} can be written in a multi-convex form:
\begin{equation}
    \label{eq:relax-multi-convex}
     \min_{\substack{\g \in [0,1]^d,\,b \in \rr,\\ \w \in \rr^{d},\,\Bu \in \rr^{d}}} \oon \sum_{i=1}^n \ell(y_i,\w^\top \C^i \g + b)+\Omega(\w) + \lambda_p\, \rho^*(\Bu)-\lambda_p \g^\top \Bu \qquad \text{s.t.}\quad {\g}^{
    \top} \mathbf{1} = s.
\end{equation}

Problem~\eqref{eq:relax-multi-convex} is convex w.r.t.~($\w$,$b$, $\Bu$) with $\g$ fixed, and with respect to 
($\g$, $b$) with ($\w$, $\Bu$) fixed. 
We therefore apply a BPCD scheme to minimize problem~\eqref{eq:relax-multi-convex}, except that, for the variable $\Bu$, the exact minimization is immediate, which we therefore use instead of a gradient update. Note that the minimization with respect to $\upsilon_j$ yields
$\upsilon_j=2 \gamma_j-1=\frac{\partial \rho}{\gamma_j}(\g)$ so that effectively the update on $\upsilon$ amounts to replacing $\rho$ by its tangent, as typically done in convex-concave optimization algorithms.
The advantage of using a multi-convex formulation is that we can use Armijo type adaptive stepsizes for each variable (In particular, the proofs of convergence in \citet{csiba2017global} generalize immediately to the case where Armijo-type linesearches are used).
Intuitively, using the tangent approximation to $\rho$ when performing a step on $\gamma$ prevents from increasing the step-size too quickly, which might help to prevent that $\gamma$ is projected ``too early'' on vertices of the capped-simplex that are suboptimal local minima. 

The algorithm that we obtain is similar to the algorithm proposed by \cite{sparsebilinear}.

\subsection{BPCD algorithm with adaptive step sizes} \label{app:BPCD-algorithm}

In this section, we detail the BPCD algorithm~\citep{xu2013block,sparsebilinear} that we use to solve the minimization problem in equation \eqref{eq:relaxed-loss}.

\paragraph{\textbf{Note:}} The objective function to minimize in problem \eqref{eq:relax-multi-convex}, which we call $\mathscr{O}$, can be decomposed in two parts, a differentiable part $h$ and a potentially non-differentiable part $\psi$ whose proximal operator can be computed efficiently:

\begin{align}
\label{eq:h}
    h(\w, \g, b, \Bu) &= \ell(y_i,\w^\top \C^i \g + b) - \lambda_p \g^\top \Bu \\
    \psi(\w, \g, b, \Bu) &= \Omega(\w) + \lambda_p \rho^*(\Bu)+\iota_{\{\g \in \Delta_s\}},
\end{align}
with $\Delta_s=\{\mathbf{\g} \in [0,1]^d \mid \mathbf{\g}^\top \ones=s\}$ the capped-simplex and $\iota_{\{\g \in \Delta_s\}}=0$ if $\g \in \Delta_s$ and $\iota_{\{\g \in \Delta_s\}}=+\infty$ else.

The main part of the BPCD algorithm is detailed in Algorithm \ref{alg:main-algo}. It consists of a main loop that involves alternative updates in $\w$, $\g$, $b$ and $\Bu$. We refer the reader to the previous section for an explanation on the dual variable $\Bu$. The loop is terminated when a convergence criterion is met.

\paragraph{\textbf{Note.}} In the following, we denote by $\p$ a set of variables $(\w, \g, \Bu, b)$. We adopt the convention that the same index $k$ indexes the set and the enclosed variables, such that: $\p^k = (\w^k, \g^k, \Bu^k, b^k)$.

\begin{minipage}{\textwidth}
\renewcommand*\footnoterule{}
\begin{algorithm}[H]
\caption{Block Proximal Coordinate Descent}\label{alg:main-algo}
\begin{algorithmic}
\State Input: $ \{\x_i, y_i\}$ for $i=1 \dots n$
\State $k \gets 0$
\State $\p^0 \gets \Call{Initialization}{ }$ \footnote{Note that when $\lambda_p>0$, ($\w^0$, $\g^0$, $\Bu^0$, $b^0$) are initialized from the previous solution.}
\State $L_{\w}, L_{\g}, L_b \gets \Call{InitializationStepsize}{\p_0}$
\State $\p \gets \p^0$
\State $k \gets 0$
\Repeat
\State $k \gets k + 1$
\State $(\w, \, L_w) \gets \Call{ProxStep\_{w}}{ \p, L_w}$ \Comment{Updating $\w$}
\State $(b, L_{b}) \gets \Call{ProxStep\_{b}}{ \p, L_{b}}$ \Comment{Updating $b$}
\State $(\g, \, L_{\g}) \gets \Call{ProxStep\_{γ}}{ \p, L_{\g}}$\Comment{Updating $\g$}
\State $\Bu \gets 2 \g - \ones$
\State $\p^k \gets p$ \Comment{Storing solution}
\Until{$\Call{ConvCrit}{\p^k, \p^{k-1}}$ \text{or} $k>\text{max\_iter}$ 
}
\State \Return $\p^k$
\end{algorithmic}

\end{algorithm}
\end{minipage}
\\

Algorithm \ref{alg:algoprox} details the proximal update step with adaptive stepsizes for the variable $\w$. It consists of a loop that searches over logarithmic decreasing stepsizes until a criterion that ensures a sufficient decrease in the objective is met. For each stepsize, the next step proposed is computed with a proximal operator. 

The updates for the variables $\g$ and $b$ are conceptually identical and can be obtained by permuting appropriately the role of the different variables.

\begin{minipage}{\textwidth}
\renewcommand*\footnoterule{}
\begin{algorithm}[H]
\caption{Proximal update step for $\w$}\label{alg:algoprox}
\begin{algorithmic}
\Function{ProxStep\_{w}}{$\p^{-}$, $L_{\text{init}}$} 
\State $m \gets -1$
\State $L \gets L_{\text{init}}$
\Repeat
\State $L \gets 
\max(L_{\min}, L \, \eta^{m})$ \footnote{$L_{\min}$ was set to $10^{-10}$, and $\eta$ to 1.5.}
\State $\w^* \gets \argminl_{\w \in \RR^d} \langle \nabla_{\w} h(\p^{-}), \w-\w^{-} \rangle + \cfrac{L}{2} \norm{\w - \w^{-}}_2^2 + \psi(\w,\g^{-},b^{-},\upsilon^{-})$
\State $\p \gets (\w^*,\g^{-},b^{-},\upsilon^{-})$
\State $m \gets m + 1$
\Until{$\Call{CritProx\_w}{\p, \p^{-}, L}$}
\State \Return $\w^*$, $L$
\EndFunction
\end{algorithmic}
\end{algorithm}
\end{minipage}
\\

The criterion for accepting the inverse stepsize $L$ is:

\begin{algorithm}[H]
\caption{Criterion stepsize for $\w$}\label{alg:criterion-stepsize}
\begin{algorithmic}
\Function{CritProx\_w}{$\p$, $\p^{-}$, $L$}
\State \Return $\big ( h(\p) < h(\p^{-}) + \langle \nabla_{\w} h(\p^{-}), \w-\w^{-} \rangle + \cfrac{L}{2} \norm{\w - \w^{-}}_2^2 \big )$
\EndFunction
\end{algorithmic}
\end{algorithm}

The BPCD algorithm terminates either when progress in the objective becomes inappreciable, or when the update in the variables are very small. The convergence criterion is formulated as: \\

\begin{minipage}{\textwidth}
\renewcommand*\footnoterule{}
\begin{algorithm}[H]
\caption{Convergence Criterion}\label{alg:convergence-criterion}
\begin{algorithmic}
\Function{ConvCrit}{$\p^k$, $\p^{k-1}$}
\State $\epsilon \gets 10^{-5}, \, \epsilon_2 \gets 10^{-10}$
\If{$\mathscr{O}(\p^{k-1}) -  \mathscr{O}(\p^k) < \epsilon \mathscr{D} \,$} 
    \State \Return true
\ElsIf{$\max(\norm{\w^k - \w^{k-1}}^2,\, (b^k - b^{k-1})^2, \norm{\g^k - \g^{k-1}}^2) < \epsilon_2 \,
$}
    \State \Return true
\Else
    \State \Return false
\EndIf
\EndFunction
\end{algorithmic}
\end{algorithm}
\end{minipage}
\\

In practice, we set the denominator $\mathscr{D}$ to $\mathscr{O}(\p^2)$ at the first relaxation iteration with $\lambda_p=0$.

\begin{algorithm}[h!]
\caption{Initialization}\label{alg:initialization}
\begin{algorithmic}
\Function{Initialization}{ }
\State $b_0 \gets 0$
\State $\w_0 \gets \mathbf{0}$
\State $\g_0 \gets \frac{s}{d} \ones$
\State $\Bu_0 \gets 2 \g - \ones$
\State \Return $\p_0$
\EndFunction
\end{algorithmic}
\end{algorithm}

The initial stepsizes are calculated with the projection of the derivative of $h$ on the Hessian.

\begin{algorithm}[h!]
\caption{Initialization of stepsize}\label{alg:initializationstepsize}
\begin{algorithmic}
\Function{InitializationStepsize}{$\p^0$}
\State $L_w \gets \cfrac{\nabla_{\w} h(\p^{0})^T \nabla^2_{\w} h(\p^{0}) \nabla_{\w} h(\p^{0})}{\norm{\nabla_{\w} h(\p^{0})}^2}$
\State $L_{\g} \gets \cfrac{\nabla_{\g} h(\p^{0})^T \nabla^2_{\g} h(\p^{0}) \nabla_{\g} h(\p^{0})}{\norm{\nabla_{\g} h(\p^{0})}^2}$
\State $L_b \gets \cfrac{\nabla_{b} h(\p^{0})^T \nabla^2_{b} h(\p^{0}) \nabla_{b} h(\p^{0})}{\norm{\nabla_{b} h(\p^{0})}^2}$ 
\State \Return $(L_{\w}^0, L_{\g}^0, L_{b}^0)$
\EndFunction
\end{algorithmic}
\end{algorithm}

\subsection{Path following algorithm and stopping criteria}
\label{app:path-following}

\paragraph{\textbf{Motivation.}} The objective function in \eqref{eq:relaxed-loss} must be solved for an increasing sequence of $\lambda_p > 0$, until $\g^* \in \{0,1\}^d$. As we noted in Section~\ref{sec:optimization-procedure}, one must be careful not to increase $\lambda_p$ too quickly. In the following, we detail the \emph{path-following algorithm} used to determine the sequence of $\lambda_p$.

\paragraph{\textbf{Rationale.}} The general procedure of the \emph{path following algorithm} consists in increasing $\lambda_p$ "progressively", each time solving problem \eqref{eq:relaxed-loss} starting from the previous solution. The criterion used to determine the next $\lambda_p$ is based on the increase in the objective $\mathscr{O}$ at the current solution. The detailed algorithm is laid out in the following paragraph.

\paragraph{\textbf{Path-following algorithm.}} First, problem \eqref{eq:relaxed-loss} is solved without push-penalty ($\lambda_p=0$), yielding a solution $\p^{(0)} = \{\w, b, \g \}$. Then,  $\lambda_p^{(1)}$ is chosen with the procedure detailed below (see equation \eqref{eq:rule-lambdaP}) and the BPCD algorithm is run again starting from $\p^{(0)}$ to minimize the objective with $\lambda_p^{(1)}$, yielding the solution $\p^{(1)}$. This procedure is repeated until $\g$ is very close to the vertices of the capped-simplex (see criterion \eqref{eq:proximity-criterion} below), or after reaching the maximum number of so-called \emph{relaxation} iterations (10000).

\paragraph{\textbf{Chosing the next $\lambda_p$.}} The rule for choosing $\lambda_p^{(i+1)}$, starting from a solution $\p^{(i)}$ is:
\begin{equation}
    \label{eq:rule-lambdaP}
    \mathscr{O}(\p^{(i)}, \lambda_p^{(i+1)}) - \mathscr{O}(\p^{(i)}, \lambda_p^{(i)}) = M \epsilon
\end{equation}
$M$ controls the tradeoff between speed and accuracy and was set to 100. $\epsilon$ is the tolerance on the objective value decrease used in the stopping criterion of the BPCD algorithm (see Alg. \ref{alg:convergence-criterion}).

\paragraph{\textbf{Stopping criterion.}} The algorithm is stopped when the solution is sufficiently close to the vertices of the capped-simplex, more precisely, when:
\begin{equation}
\label{eq:proximity-criterion}
    \sum_j{\vert \gamma_j - \tilde{ \gamma}_j \vert} < \delta,
\end{equation}
with $\tilde{\gamma}_j = \text{sign}(\gamma_j - 0.5)$ the rounded version of $\gamma_j$, and where $\delta$ was set to $10^{-10}$.

\paragraph{\textbf{Note.}} For a better coordination with the path following algorithm, the stopping criterion presented in the BPCD algorithm \cite{sparsebilinear} was changed to an absolute criterion on the loss change at the last iteration (see Algorithm \ref{alg:convergence-criterion}).

\paragraph{\textbf{References.}} The above \emph{path-following algorithm} was developed in \cite{path-following-algorithm} to solve a graph-matching problem. Here we have used a slightly simpler version than the one detailed in the paper.

\subsection{Normalization}

To limit the scaling of the inverse stepsize $L_w$ with $s$ (in algorithms \ref{alg:main-algo} and \ref{alg:algoprox}), it is useful to normalize the bilinear product $\w^\top \C^i \g$ by an appropriately chosen constant $\kappa$ which we set to $s$ for reasons now explained.

It is a classical result that a proximal step with any constant $L$ larger than the Lipschitz constant of the gradient of the function produces a valid update. In other words, any step-size of the form $\frac{1}{L}$ where $L$ is larger than the Lipschitz constant is a valid stepsize. To obtain stepsizes that have a reasonable scaling with respect to different hyperparameters such as $s$ or $d$, it can be useful to normalize the data or the loss function such that the Lipschitz constant of the gradient is controlled.
 
In this section, we thus bound $\abs{\cfrac{\partial^2{h}}{\partial{w_j^2}}}^2$ and use the fact that any continuous function with bounded derivative is also Lipschitz continuous. More precisely, the Lipschitz constant equals $2M$, $M$ being the bound on the derivative.

The bound on $\abs{\cfrac{\partial^2{h}}{\partial{w_j^2}}}^2$, where we introduce the normalization factor $\kappa$ to determine, is calculated as follows:

\begin{equation}
    \cfrac{\partial^2{h}}{\partial{w_j^2}} = \oon{\sum_{i=1}^n{\cfrac{\partial^2{\sigma}}{\partial{z_i^2}} \left( \cfrac{\partial{z_i}}{\partial{w_j}}\right)^2}}, \, \text{with} \,
    z_i = \cfrac{\w^\top \C^i \g}{\kappa} + b \,.
\end{equation}

Using the fact that $\g$ lies in $\left[0,1\right]^d$, that $|\C^i_{jk}|\leq 1,\, \forall i,j,k$ and that $\sum_j{\gamma_j} = s$, we get:

\begin{equation}
    \left(\cfrac{\partial{z_i}}{\partial{w_j}}\right)^2 = \left(\cfrac{\sum_k{\C^i_{jk} \gamma_k}}{\kappa}\right)^2
    \le \left(\cfrac{\norm{\g}_1}{\kappa}\right)^2
    \le \left(\cfrac{s}{\kappa}\right)^2.
\end{equation}

For reasons now obvious, if we set $\kappa$ to $s$, and combine the previous equations, using the fact that the second derivative of the sigmoid function is bounded by $M_{sig} = \cfrac{1}{6 \sqrt{3}}$ we get:

\begin{equation}
    \abs{\cfrac{\partial^2{h}}{\partial{w_j^2}}}^2 < M_{sig}.
\end{equation}

\section{Classifiers} \label{app:classifiers}

In this section, we describe the implementation of all classifiers used in the results section. Each classifier implemented the "balanced" class-weight setting: each observation is re-weighted, such that the weight of each class is equal in the objective. 

\subsection{SingleCellNet}
The code for SingleCellNet was downloaded from the repository at \url{https://github.com/pcahan1/singleCellNet}. In our code for the comparison of classifiers, we wrapped the original code into a scikit-learn compatible classifier.
In the first section, we detail the steps of the pipeline of SingleCellNet to select informative genes. In the second, we detail the classification algorithm, in the case where the output variable is binary (which is our case).

\subsubsection{Gene selection} \label{app:single-cell-net-app-preprocessing}

The first step of the pipeline consists in down-sampling the expression values to 1500 counts per observation and scaling it up such that the total expression per observation is 10000. Secondly, a log-transformation is applied to the expression data, and each gene is scaled to unit variance and zero mean.\\
Then, a first skimming step is applied: it retains genes which are expressed in more than $\alpha_1$ observations, or for which the average expression among observations where it is expressed (at least $\alpha_2$) is above a threshold $\mu$. After this, the correlation coefficient between the gene expression and the output variable is calculated for each gene, and the $n_{\text{nTopGenes}}$ with lowest and highest (signed) correlation coefficients are retained. \\
We use the default values of SingleCellNet, namely $\alpha_1 = .05$, $\alpha_2 = .001$, $\mu = 2$ and $n_{\text{TopGenes}} = 100$.

\subsubsection{Classification algorithm}
The expression matrix is transformed into a binary $n_{obs} \times n_{\text{TopGenes}}^2$ matrix, where each column encodes the orientation of the corresponding gene pair. The correlation coefficients between each column and the output are calculated and the $n_{\text{TopPairs}}$ top columns are retained. We use the default value of $n_{\text{TopPairs}} = 100$. The resulting matrix, supplemented by 100 random observations obtained by shuffling, is used as input to a random forest (implemented with scikit-learn) consisting of 1000 trees, trained to optimized the balanced accuracy. The other parameters of the random forest are set to default values (see Appendix \ref{app:random-forest} for a description of the default values).

\subsection{Random Forest} \label{app:random-forest}
The classifier \rf ~ was implemented with the scikit-learn library, setting the number of trees to $300$ and the class-weight to "balanced". We set the number of trees to $300$ after verifying empirically that a larger number of trees wouldn't produce better solutions. The other hyperparameters were set to default values. Namely, each split minimizes the "Gini" impurity of resulting leaves, each leaf is split until it is pure, $\sqrt{d}$ randomly chosen variables are considered at each split, and $n_{obs}$ observations are drawn (with repetition) to train each tree.

\subsection{Logistic Regression} \label{app:logistic-regression}
The classifier \lr ~ was implemented with the scikit-learn library, specifying the maximum number of iterations to 10 000, the tolerance to $10^{-3}$, choosing the \textsc{saga} solver \citep{saga}, and setting the class-weight to "balanced". Regularization was not applied if not specified (Note that to achieve zero $\ell_2$-regularization, one must set the $C$ parameter to a high value).

\subsection{Logistic Regression on ranks} \label{app:ranklr}
A rank-transformation (with \emph{average ranking}, see Appendix \ref{sec:ties-handling}) was applied to the data prior to the logistic regression classifier detailed above in Appendix \ref{app:logistic-regression}.

\subsection{Optirank}
\textbf{Implementation.} The source code of our implementation of \optirank~ is available on the repository \url{https://github.com/paolamalsot/optirank}. The algorithm is detailed in Appendices \ref{app:BPCD-algorithm} and \ref{app:path-following}.

\paragraph{\textbf{Note.}} For the classification tasks on real datasets (in Section~\ref{sec:experiments-real-data}), \lr~was used to refit $\w$ and $b$ with frozen $\g$. The goal is to prevent the incompatible stopping criteria between \lr ~ and \optirank ~ to interfere with the comparison between both classifiers.

\textbf{Balanced setting.} As all the other classifiers, \optirank ~ was trained with data points reweighted inversely proportionally to the size of each class, to balance the classes.

\subsection{ANrank-lr} \label{app:ANranklr}
The classifier is similar to \ranklr ~ (detailed previously in app. \ref{app:ranklr}), except that the ranking transformation is applied with a previously chosen \emph{ranking reference set}. The \emph{ranking reference set} is chosen with a simple ANOVA that eliminates from the reference set genes whose expression is subject to dataset shift. Note that \ANranklr ~ uses the \emph{auxiliary source dataset} only for the ranking reference set selection. For each gene, a two-way ANOVA with factors \emph{label} and \emph{dataset} is carried out. We then select as ranking reference the genes which have the smallest F-value corresponding to the marginal effect of dataset - in other words, the ones less susceptible to dataset-shift.

\section{Synthetic example: Supplementals} \label{sec:simulation-supp}
\subsection{Generation of \texorpdfstring{$\w$, $\g$ and $b$}{w, γ and b}}
In this section, we describe the generation of parameters $\w$, $\g$ and $b$ for the simulation of the task described in Section~\ref{sec:toy-model}.

Let us first recall from equation \ref{eq:simulation-y} that the label of each sample follows a simple logistic model on the ranks within the stable genes: 

\begin{align}
    \PP(Y=1|X=\x_i)&=\sigma(\w^\top \r_i + b) \quad \text{with} \: 
    & w_j=0 \; \forall j \notin S, \;\
    & \Gamma = S. \\
    &=\sigma(\w^\top \mathbf{C}^i \g + b).
\end{align}

In practice, once $\w$, $\g$ and $b$ are generated, it suffices to draw the label from a Bernoulli distribution whose probability of \emph{success} is given by the above equation. 

\paragraph{Simulation procedure.} First, the design matrix $\mathbf{X}$ is generated following the model detailed in Section~\ref{sec:toy-model}. From the synthetic model, $\g$ is fixed to $\g_i = 1 \, \forall i \in S, \, 0 \, \text{otherwise}$. A non-scaled version of $\w$ is sampled as: $w_i \sim (-1)^\text{sgn} (\mathcal{N}(0,1) + 1)$, with $\text{sgn} \sim \mathcal{B}(0.5)$. b is then adjusted to ensure that the labels generated are balanced (i.e. that the average of the probability of generating a positive class is 0.5 across observations). Next, $\w$ and $b$ are scaled by the same factor to control the noise of the generated data. The noise is such that in expectation, the balanced accuracy between the most probable label and the label drawn is 98 \%.

\subsection{Classifiers comparison} \label{app:simulation-details}

\paragraph{\textbf{Simulation parameters.}} The default parameters for the simulation task were set to: $d=50$, $d_P=40$, $n = 1000$, $\tau = 0.2$, and $\sigma = 0.05$. Figures \ref{fig:simulation-main} and \ref{fig:simulation-supp} indicate the score for varying values of simulation parameters ($d \in \{50, 100, 150, 200 \}$, $d_P \in \{0, 10, 20,30, 35, 40\}$, and $\tau \in \{0, 0.125, 0.25, 0.375, 0.5\}$). Each simulation experiment was repeated 4 times, resulting in the error-bars displayed on the figures.

\paragraph{\textbf{Note.}} In the experiment with increasing values for $d$, we also increased the size of the training set $n$ to keep the ratio $n/d$ constant.

\paragraph{\textbf{Cross-validation.}} The generated dataset was split in a stratified fashion, keeping $30\%$ for the test. Each classifier was trained and tuned with 5-fold internal cross-validation on the train set.

\paragraph{\textbf{Hyperparameter selection.}} The parameter grid for classifiers \lr~and \ranklr~consisted in 5 log-spaced values for the $\ell_2$ regularization (0, 0.0001, 0.001, 0.01, 0.1), corresponding to the parameter $\lambda_2$ in $\Omega(\w)$ (see section \ref{sec:classifiers}). In addition, the grid for \optirank~included 5 values for the hyperparameter $s/d$ $(0.2, 0.4, 0.6, 0.8, 1)$. The hyperparameter $\lambda_1$ was set to 0. For each classifier, we selected the model with highest validation balanced accuracy.

\subsection{Supplementary figures} \label{app:supp-figure}

\subsubsection{Simulation results for varying \texorpdfstring{$\tau$}{τ}}

\paragraph{\textbf{Motivation.}} We investigated how the outcome of the comparison between the competing classifiers (\lr, \ranklr~and \optirank) is affected by the \emph{simulation parameter} $\tau$. $\tau$ defines the magnitude of the coordinated shift of the perturbed genes in $P$ from one observation to the other.

\begin{figure}[H]
\centering
\includegraphics[width=0.75\linewidth]{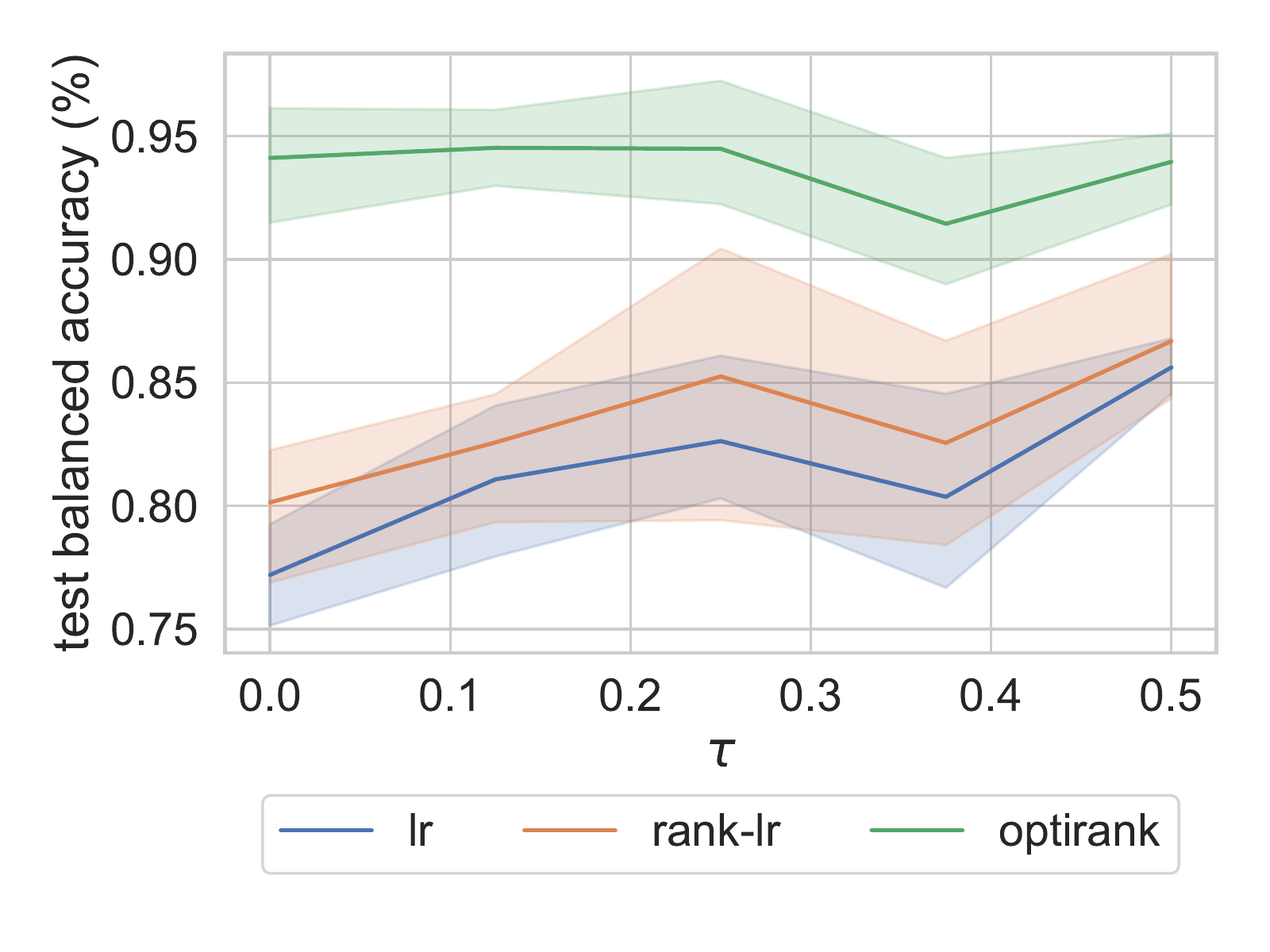}
\caption{Dependence of test-accuracy with respect to the simulation parameter $\tau$. Default simulation parameters were set to $d=50$, $d_P=40$, $n = 1000$, and $\sigma = 0.05$.}
\label{fig:simulation-supp}
\end{figure}

\paragraph{\textbf{Results.}} Figure \ref{fig:simulation-supp} shows that the superiority of \optirank~over \ranklr~and \lr~is maintained throughout the \emph{simulation parameter} range that we explored.

\subsubsection{Overlap between the true \texorpdfstring{$\gamma$}{γ} and the one fitted by \optirank}

\begin{figure}[h!]
\begin{multicols}{2}
\centering
\includegraphics[width=\linewidth]{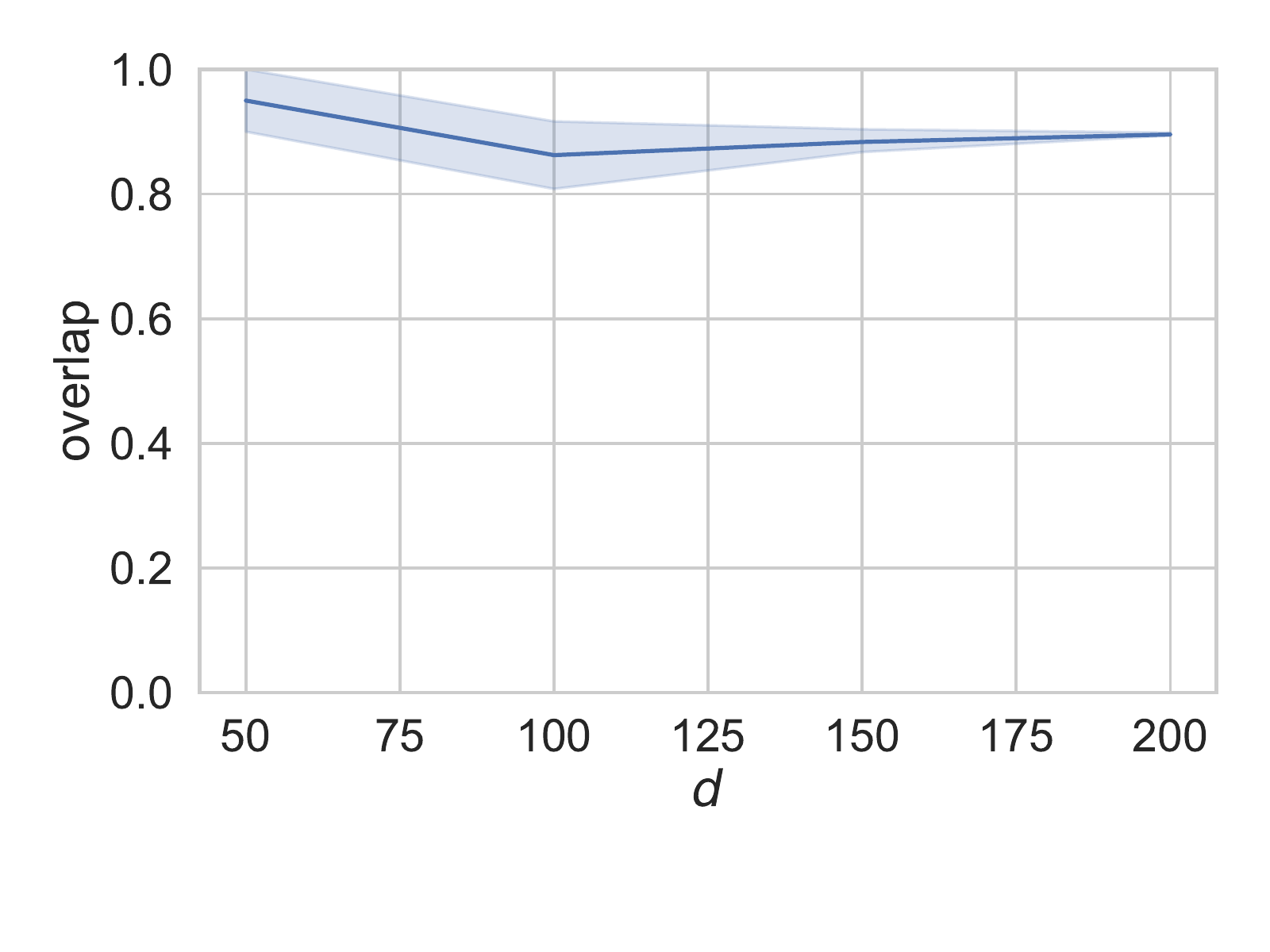}
\\
\includegraphics[width=\linewidth]{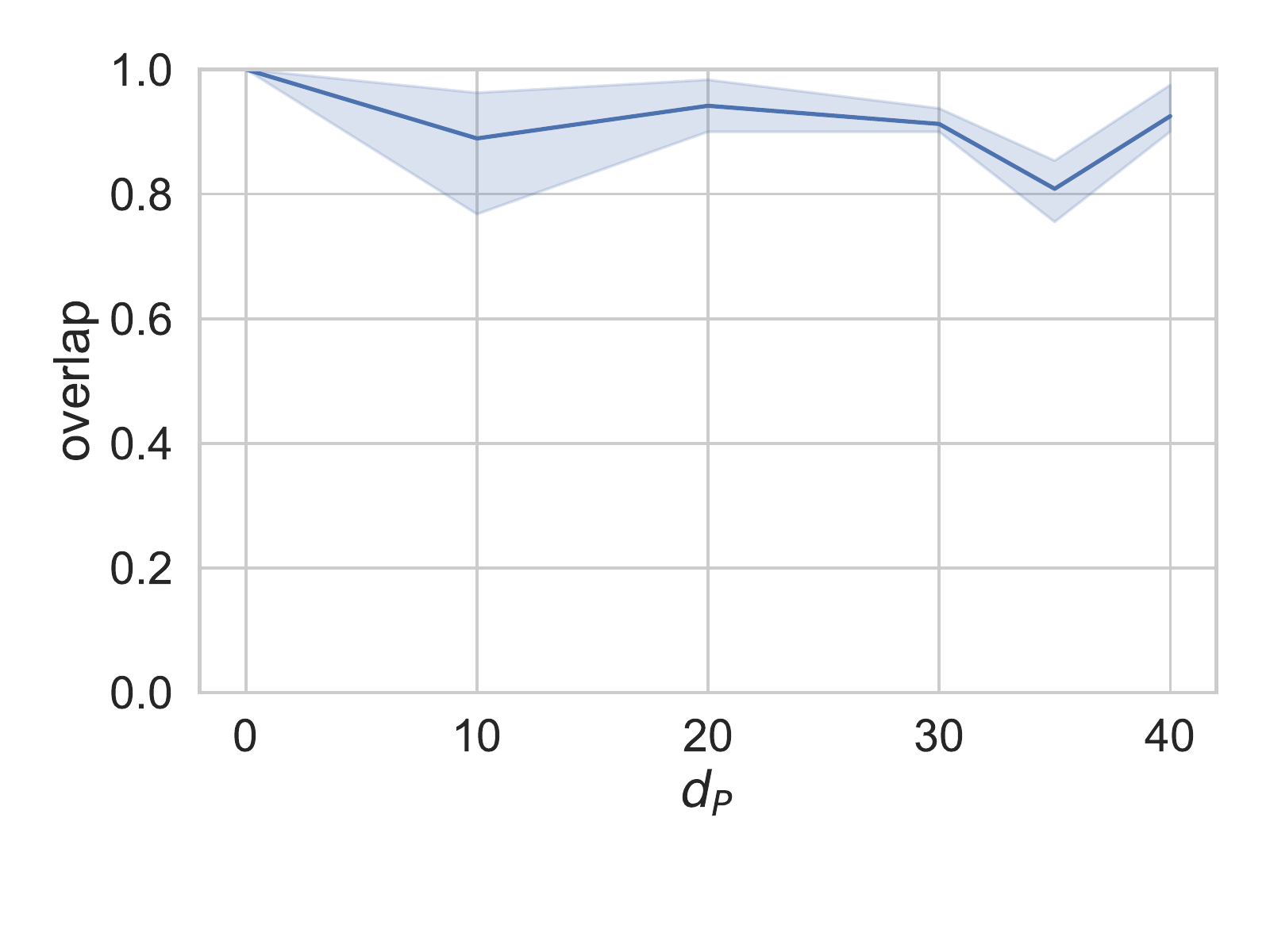}
\end{multicols}
\caption[short]{Cosine similarity $C_S$ between the true $\gamma$ and the one fitted by \optirank, as a function of the dimension $d$ and the number of perturbed genes $d_P$. The shaded area shows the standard error across runs. Default simulation parameters were set to $d=50$, $d_P=40$, $n = 1000$, $\tau = 0.2$, and $\sigma = 0.05$.}
\label{fig:simulation-overlap}
\end{figure}

\clearpage

\section{Results on real data} \label{app:results-on-real-data}

\subsection{Datasets} \label{app:datasets}

In this section we detail the data sources and the selection of examples used for each of the classification tasks.

\paragraph{\textbf{TCGA.}} The public count-table for the RNA-Seq data (with dbGab accession number phs000178.v10.p8) and the metadata was downloaded from the GDC data portal (\url{https://portal.gdc.cancer.gov}). We selected primary tumors with RNA-Seq data, with disease type in ('Adenomas and Adenocarcinomas', 'Squamous Cell Neoplasms', 'Cystic, Mucinous and Serous Neoplasms', 'Ductal and Lobular Neoplasms') and tissue of primary in ('Breast','Bronchus and lung','Esophagus','Stomach','Colon','Stomach','Pancreas','Rectum', 'Rectosigmoid junction', 'Prostate'). The label for the classification task was taken from the tissue of primary, by agglomerating "Colon", "Rectum" and "Rectosigmoid junction" in one class called "Colorectal".

\paragraph{\textbf{Note.}} In the task \task{TCGA-PCAWG-met500}, we ensured the binary problems One-vs-Rest involved a positive class present in the \emph{auxiliary source} dataset PCAWG. For this reason, we had to remove the binary problems "Bronchus and lung" VS Rest and "Prostate" VS Rest.

\paragraph{\textbf{PCAWG.}} The PCAWG dataset comprises among other the TCGA dataset. Care was taken to exclude instances from the PCAWG that belong to the TCGA dataset during the selection of examples. The RNA-Seq data was downloaded from \url{https://dcc.icgc.org/releases/PCAWG/transcriptome/transcript_expression} and the metadata from \url{https://dcc.icgc.org/releases/PCAWG/transcriptome/metadata}. The same disease types and tissues of primary as in TCGA were selected, the class labels were processed in the same way.

\paragraph{\textbf{BRCA.}} The mutation status of the BRCA1 and BRCA2 genes for the examples selected in the TCGA was obtained on the GDC data portal, and aggregated into one binary class label indicating the presence of at least one mutation among BRCA1 and BRCA2. We selected instances whose tissue of primary was the Breast. The expression data is the same as the one in the TCGA.

\paragraph{\textbf{met-500.}} The expression data and the metadata was downloaded from \url{https://xenabrowser.net/datapages/?cohort=MET500\%20(expression\%20centric)}. Only metastatic tumors were selected, of the same primary tissue as in the TCGA. The class-label was generated with the tissue of primary.

\paragraph{\textbf{Single-Cell datasets: Baron, Murano, Segerstolpe, MWS, TM10x, TMfacs}}
The datasets were downloaded from \url{https://github.com/pcahan1/singleCellNet#trainsets}. For each task consisting of a pair/triplet of datasets, only common genes and common cell types were selected. Here is a short description per dataset:

\begin{description}[labelindent=1cm]
    \item \textbf{Murano.} Adult human pancreatic cells, sequenced with CEL-Seq2 (2120 cells)
    \item \textbf{Baron.} Adult human pancreatic cells, sequenced with inDrop (8596 cells)
    \item \textbf{Segerstolpe.} Adult human pancreatic cells, sequenced with Smart-Seq2 (2209 cells)
    \item \textbf{MWS.} Adult mouse cells, across 125 cell types, sequenced with Microwell-seq (6477 cells)
    \item \textbf{TM10x.} Adult mouse cells, across 32 cell types, sequenced by 10x (1599 cells)
    \item \textbf{TMfacs.} Adult mouse cells across 69 cell types, sequenced by smartseq2 (3182 cells)
\end{description}

\subsection{Pre-processing} \label{app:datasets-preprocessing}
All datasets were pre-processed with the pre-processing pipeline of SingleCellNet, detailed in Appendix \ref{app:single-cell-net-app-preprocessing}.

\subsection{Cross-validation splits}

In this appendix, we detail, per experiment and for each classification task, which dataset(s) and which training-testing scheme were used. All splits were realized in a stratified fashion.

\subsubsection{Cross-validation splits for the tasks of section \ref{sec:experiment1}.} \label{app:datasets-cv1}

\paragraph{\textbf{\task{TCGA.}}} 10\% of the TCGA dataset was held out as a test set. The remaining 90\% was used for 5-fold internal cross-validation.

\paragraph{\textbf{\task{PCAWG.}}} The same training set and cross-validation splits as in \task{TCGA} were used. The PCAWG dataset was used as a test set.

\paragraph{\textbf{\task{met-500.}}} The same training set and cross-validation splits as in \task{TCGA} were used. The met-500 dataset was used as a test set.

\paragraph{\textbf{\task{BRCA.}}} The dataset was split in four equal parts (each containing the same percentage of BRCA1 and BRCA2 mutated instances). A nested-cross validation scheme was applied in which for every possible combination, 2 parts are used for training, the third for validation, and the fourth for testing.

\paragraph{\textbf{\task{Cell-typing single-cell data.}}} For each task called X-Y (for instance Baron-Murano), dataset X was used as a training set and dataset Y as a test set.
The training data was generated by randomly picking (maximum) 100 cells for each cell type. The rest of the training data is used for validation. This process is done 5 times. The test dataset is used as a whole.

\subsubsection{Cross-validation splits for the tasks of section \ref{sec:multi-source training} in the \emph{multi-source} and \emph{single-source} scenario}\label{app:CV-multi-source}

We now describe the cross-validation splits in the tasks involving three datasets: one main \emph{source} dataset, a \emph{auxiliary source} dataset and a \emph{target} dataset.
These tasks are named with the pattern X-Y-Z, where X is the main \emph{source} dataset, Y the \emph{auxiliary source} dataset and Z the \emph{target} dataset (for instance TCGA-PCAWG-met500). we distinguish two training scenarios: the \emph{multi-source} and \emph{single-source} scenarios. In the \emph{multi-source} scenario, datasets X and Y are merged, both in the train and validation splits. \\ In the \emph{single-source} scenario dataset Y is not used (except by \ANranklr~which uses dataset Y in its entirety, for the sole purpose of selecting the reference genes during the training phase). Indeed, in the \emph{single-source} scenario, \ANranklr~does not use dataset Y for fitting the coefficients of the logistic regression nor for validation.

\paragraph{\textbf{\task{TCGA-PCAWG-met500.}}} We used 5-fold cross-validation to divide the TCGA and PCAWG datasets into train/validation splits, for each dataset separately. In the \emph{single-source} scenario, only the TCGA is used. In the \emph{multi-source} scenario, we compose each train and validation split by doing the union of the corresponding splits in the TCGA and PCAWG cross-validation splits.

\paragraph{\textbf{\task{Cell-typing single-cell data.}}} In the \emph{single-source} setting, we pick randomly from dataset X 100 cells for each cell-type, the rest of dataset X goes in the validation split. We repeat this 5 times to generate 5 validation folds. \ANranklr ~ uses a small(er) version of dataset Y, which comprise 100 cells for each cell-type. \\
In the \emph{multi-source} scenario, we took care not to augment the size of the training set. To achieve this, we used the same CV splits as in the \emph{single-source} setting, except that we downsampled to 50 the number of examples from dataset X for each cell-type, to which we added 50 examples of dataset Y for each cell-type. We added to the validation splits generated in the \emph{single-source} setting the portion of dataset Y which was not used in the training set.

\subsection{Hyperparameter grid} \label{app:results-hp-grid}

In table \ref{tab:results-hyper-params}, we list the hyperparameters tuned for each classifier, and the corresponding values which were tried in the parameter grid.

\begin{table}[H]
    \centering
    \begin{tabular}{|l|c|c|}
    \toprule
       Classifier & Parameter Name & Values \\
       \midrule
       \optirank & $\lambda_1$ & \{0, 0.0001, 0.001, 0.01, 0.1\} \\
        & $\lambda_2$ & \{0, 0.0001, 0.001, 0.01, 0.1\} \\
        & $\nicefrac{s}{d}$ & \{0.005, 0.01, 0.02, 0.05, 0.1, 0.2, 0.4, 0.6, 0.8, 1.0\} \\
        \hline
        \ANranklr & $\lambda_1$ & \{0, 0.0001, 0.001, 0.01, 0.1\} \\
        & $\lambda_2$ & \{0, 0.0001, 0.001, 0.01, 0.1\} \\
        & $\nicefrac{s}{d}$ & \{0.005, 0.01, 0.02, 0.05, 0.1, 0.2, 0.4, 0.6, 0.8, 1.0\} \\
        \hline
        \lr & $\lambda_1$ & \{0, 0.0001, 0.001, 0.01, 0.1) \\
        & $\lambda_2$ & \{0, 0.0001, 0.001, 0.01, 0.1\} \\
        \hline
        \ranklr & $\lambda_1$ & \{0, 0.0001, 0.001, 0.01, 0.1\} \\
        & $\lambda_2$ & \{0, 0.0001, 0.001, 0.01, 0.1\} \\
        \hline
        \rf & - & - \\
        \hline
        \SCN & - & - \\
        \bottomrule
    \end{tabular}\\[2mm]
    \caption{Hyperparameters for classifiers in Section~\ref{sec: results-and-discussion}. Note that the dimension $d$ is 1000 for all classifiers.}
    \label{tab:results-hyper-params}
\end{table}

Note that in the scikit-learn implementation of the logistic regression, the elastic net regularization is prescribed by the parameters $C$ and $\ell_1$\_ratio. Therefore, these parameters were set according to the values of $\lambda_1$ and $\lambda_2$.

\subsection{Comparison between classifiers} \label{app:comparison-between-classifiers}

\paragraph{\textbf{Scoring function: Balanced Accuracy.}}

The balanced accuracy scoring function was used as a a scoring metric both for hyperparameter selection and for model evaluation. \\
The balanced accuracy equals the average of sensitivity (true positive rate) and specificity (true negative rate), and is defined as such:

 \begin{equation}
     \text{balanced accuracy} = \nicefrac{1}{2}\left(\cfrac{TP}{TP + FN} + \cfrac{TN}{TN + FP}\right),
 \end{equation}
 
 where TP stands for the number of true positives, FN for the false negatives, etc..
 
 Note that in a balanced setting, the above formula reduces to the conventional accuracy. By contrast, in an unbalanced setting, a classifier which predicts all the time the predominant class will achieve a score of $\nicefrac{1}{2}$.
 
 \subsubsection{Hyperparameter selection and model evaluation}
 
 In this subsection, we describe how hyperparameters were chosen and how the resulting model was evaluated and compared to its competitors.
 
 Each multi-class classification task was separated in $n_{classes}$ One-VS-Rest binary classification tasks, the binary classification tasks were left as is. Each classifier was trained for each parameter combination, in each binary problem, for each (internal) training split, and tested in the corresponding validation set and on the test set. In each binary task, for every test set (there is a unique one apart in the task \task{BRCA}), we selected the hyperparameters with the one-standard-error rule applied to the balanced accuracy averaged over validation splits. More precisely, we selected the sparsest model whose averaged balanced accuracy was within one standard error of the highest averaged balanced accuracy. We did not refit each classifier with the optimal hyper-parameter combination on the whole training set. Instead, for computational reasons, we kept as many models as the number of validation splits, and, as a result, the test-score, as displayed in table \ref{tab:summary-results-simple_tasks}, was calculated with the average test score among those classifiers.

\subsubsection{Pairwise comparisons}
\label{sec:pairwise-comparisons}

The following section details how each pair of classifier was compared, in order to determine if one performs \emph{significantly} better than the other. In order to do so, we performed the comparison per dataset. We computed the difference in balanced accuracy between the two classifiers for each validation fold and each binary classification task (in such a way that no average is involved). Every one of those differences was accumulated as an independent observation, and a paired Student's t-test was used to determine, based on the differences, if one classifier outperformed the other on the dataset at hand. (We used a two-sided test with a significance level of 5 \%). \\

Table \ref{table:pairwise-BRCA} through \ref{table:pairwise-TCGA_PCAWG_met500} show the result of the pairwise comparisons in each task. On the lower left part of each table, the sign at position (i,j) indicates if the classifier corresponding to row i is significantly better (+) or worse (-) than the classifier in column j. The percentages in the upper-right part indicate the fraction of instances where classifier i performed better than classifier j. In addition, whenever this percentage is associated with the algorithm on line i significantly outperforming the algorithm in column j the number is highlighted in bold green, and conversely, if this number correspond to a case where the algorithm on row i performs significantly worse then algorithm in column j, it is highlighted in red italic.

\subsubsection{Pairwise comparisons for the classification tasks of section \ref{sec:experiment1}}

\makeatletter
\@for\sun:={BRCA,TCGA,PCAWG,met-500,Baron_Murano,Baron_Segerstolpe,MWS_TM10x,MWS_TMfacs,TM10x_MWS,TM10x_TMfacs,TMfacs_MWS}\do{\input{tables/pairwise_comparisons/simple_tasks/\sun.txt}}
\makeatother

\clearpage

\subsubsection{Pairwise comparisons for the classification tasks of section \ref{sec:multi-source training}, in the \emph{single-source} scenario}

\makeatletter
\@for\sun:={Baron_Segerstolpe_Murano,MWS_TMfacs_TM10x,TCGA_PCAWG_met500}\do{\input{tables/pairwise_comparisons/single_source/\sun.txt}}
\makeatother

\clearpage

\subsubsection{Pairwise comparisons for the classification tasks of section \ref{sec:multi-source training}, in the \emph{multi-source} scenario}

\makeatletter
\@for\sun:={Baron_Segerstolpe_Murano,MWS_TMfacs_TM10x,TCGA_PCAWG_met500}\do{\input{tables/pairwise_comparisons/multi_sources/\sun_01_sub_merged.txt}}
\makeatother

\clearpage

\subsubsection{Comparisons in terms of sparsity}\label{sec:sparsity-investigation}

In Section~\ref{sec: results-and-discussion}, we highlighted the fact that \optirank, compared to \ranklr~has the added beneficial potential to produce sparse solutions. In fact, by definition, \ranklr~necessitates to know the value of all genes to perform the ranking. \\ However, it was mentioned that on instances where \lr~performed well, there was no advantage in using \optirank, as the solutions of \lr~are usually sparser than the ones produced by \optirank. In the following, we support this claim. 

As a first investigation, we computed per dataset the average number of genes of the solution found by \optirank~and \lr. Note that for \optirank, the effective number of genes is the number of genes which are either in the reference set indicated by $\g$ or whose corresponding coefficient $w_j$ is non-zero. Table \ref{tab:average-number-of-genes} summarizes the results and it is clear that \lr~produces sparser solutions than \optirank.

\hspace*{-4cm}
\begin{table}[H]
\centering

\begin{tabular}{|l|cccc|}
\toprule
{} &  $\overline{\sorlr}$ &  $\overline{\slr}$ &  $100\,\hat{P}(\sorlr\!\le\!\slr)$ &  $100\,\hat{P}(\slr\!\le\!\sorlr)$ \\
\midrule
\task{BRCA}              &       454 &      43 &                                      8 &                                   92 \\
\task{TCGA}              &       598 &     588 &                                     24 &                                   76 \\
\hline
\task{PCAWG}             &       598 &     588 &                                     24 &                                   76 \\
\hline
\task{met-500}           &       598 &     588 &                                     24 &                                   76 \\
\hline
\task{Baron-Murano}      &       462 &     406 &                                     57 &                                   42 \\
\task{Baron-Segerstolpe} &       511 &     451 &                                     44 &                                   67 \\
\task{MWS-TM10x}         &       629 &     365 &                                     21 &                                   79 \\
\task{MWS-TMfacs}        &       666 &     226 &                                      9 &                                   91 \\
\task{TM10x-MWS}         &       514 &     242 &                                     27 &                                   73 \\
\task{TM10x-TMfacs}      &       616 &     305 &                                     36 &                                   64 \\
\task{TMfacs-MWS}        &       530 &     204 &                                     25 &                                   75 \\
\bottomrule
\end{tabular}\\[2mm]

\caption{Number of genes $\sorlr$ and $\slr$ in the solutions produced by \optirank~and \lr ~ (2 outermost left columns), averaged across folds and classes. The third column shows the percentage $100\,\hat{P}(\sorlr\!\le\!\slr)$ of instances where the solution of \optirank~is sparser than the solution found by \lr. The last column shows the percentage $100\,\hat{P}(\slr\!\le\!\sorlr)$ of instances with the opposite case, i.e where the solution of \optirank~is less sparse than the solution found by \lr.}
\label{tab:average-number-of-genes}
\end{table}

In fact, in the task \task{BRCA}, \optirank~rarely produces a sparser solution, and on other tasks, apart from the task \task{Baron-Murano} and \task{TMfacs-MWS}, \lr~produces sparser solutions in a majority of cases. It should be clear however that in all cases where some robustness can be obtained by using a rank representation, the classical rank representation requires to measure all genes and has thus no sparsity whatsoever, while \optirank~attempts by construction to use a reduced set of genes, and so, even if the level of sparsity obtained is not comparable to that of an \lr~model the gain in performance might be worth it.

\subsubsection{Comparisons in terms of sparsity in the \emph{multi-source} scenario}\label{sec:sparsity-investigation-multi-source}

We carried the same analysis than in the previous section for the tasks in the \emph{multi-source} scenario (see Section \ref{sec:multi-source training}). As in previous section, \lr~produces sparser solutions than \optirank~and \ANranklr~in a majority of cases. It is worth noting that solutions found by \ANranklr~and \optirank~require a similar number of genes, with \optirank~producing slightly sparser solutions.

\begin{table}[H]
\addtolength{\leftskip} {-3cm} 
\addtolength{\rightskip}{-3cm}
\centering
\resizebox{\columnwidth}{!}{%
\begin{tabular}{|l|cccccc|}
\toprule
{} &  $\overline{\sorlr}$ &  $\overline{\sAr}$ &  $\overline{\slr}$ &  $100\,\hat{P}(\sorlr\!\le\!\sothers)$ &  $100\,\hat{P}(\sAr\!\le\!\sothers)$ &  $100\,\hat{P}(\slr\!\le\!\sothers)$ \\
\midrule
\task{TCGA-PCAWG-met500}        &       844 &     849 &     273 &                                     20 &                                    0 &                                   80 \\
\task{Baron-Segerstolpe-Murano} &       606 &     658 &     364 &                                     12 &                                   12 &                                   75 \\
\task{MWS-TMfacs-TM10x}         &       604 &     796 &     300 &                                     15 &                                    2 &                                   83 \\
\bottomrule
\end{tabular}}\\[2mm]
\caption{Number of genes $\sorlr$, $\sAr$, and $\slr$ in the solutions produced by \optirank, \ANranklr~and \lr~(3 outermost left columns), averaged across folds and classes. The last three columns shows the percentage of instances where the solution of the indicated classifier is sparser than the ones found by the other classifiers.}
\label{tab:average-number-of-genes-multi-source}
\end{table}

\subsubsection{Supplementary table for \emph{single-source} scenario} \label{app:results-single-source}

\begingroup
\setlength{\tabcolsep}{2pt}
\begin{table}[h!]
\centering
\addtolength{\leftskip} {-3cm} 
\addtolength{\rightskip}{-3cm}
\label{tab:summary-results-single_source}
\resizebox{\columnwidth}{!}{%
\begin{tabular}{|l|cccccc|}
\toprule
{} &                         \ANranklr &                \SCN &                                   \lr &                             \optirank &                               \ranklr &             \rf \\
\midrule
\task{TCGA-PCAWG-met500}        &  $\mathbf{80 \pm 3}$ \textbf{(2)} &      $66 \pm 4$ (5) &                        $74 \pm 4$ (4) &      $\mathbf{81 \pm 4}$ \textbf{(1)} &      $\mathbf{76 \pm 3}$ \textbf{(3)} &  $61 \pm 4$ (6) \\
\task{Baron-Segerstolpe-Murano} &  $\mathbf{93 \pm 1}$ \textbf{(3)} &  $89.4 \pm 1.5$ (5) &                    $89.5 \pm 2.4$ (4) &  $\mathbf{93.7 \pm 1.5}$ \textbf{(2)} &  $\mathbf{93.8 \pm 1.7}$ \textbf{(1)} &  $62 \pm 3$ (6) \\
\task{MWS-TMfacs-TM10x}         &  $\mathbf{82 \pm 2}$ \textbf{(4)} &      $72 \pm 2$ (5) &  $\mathbf{83.3 \pm 2.2}$ \textbf{(2)} &      $\mathbf{84 \pm 2}$ \textbf{(1)} &                    $83.0 \pm 2.4$ (3) &  $53 \pm 1$ (6) \\
\bottomrule
\end{tabular}}\\[2mm]
\caption{\textbf{\emph{Single-source scenario}.} Average balanced accuracies in \% (across folds and classes) of competing classifiers on the tasks detailed in section~\ref{sec:multi-source training} in the case in which the \emph{auxiliary source} dataset is not used (except for \ANranklr ~ which uses it for ranking reference genes selection). The integer in parenthesis denotes the rank of the classifiers in terms of average balanced accuracy (lower is better). Classifiers which did not score significantly worse than the best classifier according to a paired Student's t-test (with a 5\% significance level) are highlighted in bold (see Appendix \ref{app:comparison-between-classifiers} for additional details).}
\end{table}

\endgroup

\subsubsection{Runtime comparisons} \label{app:runtime-estimates}
Table \ref{tab:runtime-estimates} shows the average runtime (in seconds) across validation splits and binary problems of competing classifiers with the hyperparameters set to their optimal values according to the one-standard-error-rule. We applied our comparison on the tasks detailed in \ref{sec:multi-source training}, with classifiers trained in the \emph{single-source} scenario. The results attest that the fitting time of \optirank~ is reasonable, and on some tasks, it is even lower than the fitting time of its competitors using rank features such as \ANranklr~and \ranklr.

\begin{table}[h!]
\centering
\addtolength{\leftskip} {-2.5cm} 
\addtolength{\rightskip}{-2.5cm}
\resizebox{\columnwidth}{!}{%
\begin{tabular}{|l|cccccc|}
\toprule
{} &  \ANranklr &          \SCN &        \lr &  \optirank &    \ranklr &          \rf \\
\midrule
\task{TCGA-PCAWG-met500}        &   135 ± 38 &   205.6 ± 0.8 &   140 ± 47 &     70 ± 4 &   150 ± 46 &    8.6 ± 0.5 \\
\task{Baron-Segerstolpe-Murano} &     10 ± 1 &    50.8 ± 0.1 &  2.0 ± 0.3 &  7.7 ± 0.7 &  3.6 ± 0.3 &  0.75 ± 0.01 \\
\task{MWS-TMfacs-TM10x}         &  8.8 ± 0.5 &  33.82 ± 0.09 &  2.5 ± 0.2 &     16 ± 2 &     36 ± 9 &  0.92 ± 0.02 \\
\bottomrule
\end{tabular}}\\[2mm]
\caption{\textbf{Runtime estimates.} Average runtime (in seconds) across validation splits and binary problems of competing classifiers for the hyperparameter set which was selected as optimal according to the one-standard-error-rule. All classifiers were trained in the \emph{single-source} scenario.}
\label{tab:runtime-estimates}
\end{table}

\section{Computational resources} \label{app:computational-resources}
The computational results in this work were produced with an internal computing cluster, in a batched fashion, amounting to a total of 2500 CPU hours. Per batch, the maximal capacity used was 30 GB of RAM with 2 CPUs.

\end{document}